\documentclass[11pt,a4paper,DIV=11,numbers=noenddot]{scrartcl}
\usepackage[utf8]{inputenc}
\makeatletter
\DeclareOldFontCommand{\rm}{\normalfont\rmfamily}{\mathrm}
\DeclareOldFontCommand{\sf}{\normalfont\sffamily}{\mathsf}
\DeclareOldFontCommand{\tt}{\normalfont\ttfamily}{\mathtt}
\DeclareOldFontCommand{\bf}{\normalfont\bfseries}{\mathbf}
\DeclareOldFontCommand{\it}{\normalfont\itshape}{\mathit}
\DeclareOldFontCommand{\sl}{\normalfont\slshape}{\@nomath\sl}
\DeclareOldFontCommand{\sc}{\normalfont\scshape}{\@nomath\sc}
\makeatother
\usepackage{amsmath,amssymb,graphicx,scalefnt,enumitem}
\usepackage[absolute]{textpos}
\usepackage{cite}
\usepackage[affil-it]{authblk}
\usepackage{xcolor}
\usepackage{tabularx,booktabs}
\usepackage{xspace}
\usepackage{listings}
\usepackage[final]{showkeys}
\usepackage{filemod}\usepackage[font=small,labelfont=bf,format=plain,margin=0.05\textwidth]{caption}
\usepackage{subfig}
\usepackage[pdftitle={The light CP-even MSSM Higgs mass resummed to fourth logarithmic order},
  pdfauthor={Robert V. Harlander, Jonas Klappert, Daniel Ochoa, Alexander Voigt},
  pdfkeywords={Himalaya,H3m,FlexibleSUSY,MSSM,Higgs,supersymmetry},
  bookmarks=true, linktocpage,
  colorlinks=true, allbordercolors=white, allcolors=blue]{hyperref}
\newcommand{\RHheaderline}{{\footnotesize TTK--18--22\\ August 2018}}
\newcommand{\fo}{\text{\abbrev{FO}}}

\newcommand{\mhiggs}{M_h}
\newcommand{\himalaya}{\code{Himalaya}}
\newcommand{\lambdabar}{\bar\lambda}
\newcommand{\mtbar}{\ensuremath{\bar m}_t}
\newcommand{\vbar}{\ensuremath{\bar v}}

\newcommand{\as}{a_s}

\newcommand{\asbar}{\bar{a}_s}

\newcommand{\mep}{m_{\ep}}
\newcommand{\htm}{\code{H3m}}
\newcommand{\one}{one}
\newcommand{\two}{two}
\newcommand{\three}{three}
\newcommand{\four}{four}
\newcommand{\lmut}{l_{\mu t}}
\newcommand{\barlmut}{\bar{l}_{\mu t}}

\newcommand{\lts}{l_{tS}}
\newcommand{\lst}{l_{St}}
\newcommand{\noeqn}[1]{(\ref{#1})}
\newcommand{\eft}{\abbrev{EFT}}
\newcommand{\ms}{\ensuremath{M_S}}

\newcounter{notecount}

\newcommand{\citere}[1]{Ref.~\cite{#1}}
\newcommand{\citeres}[1]{Refs.~\cite{#1}}
\newcommand{\code}[1]{\texttt{#1}}
\newcommand{\abbrev}[1]{{\scalefont{.9}#1}}

\newcommand{\ep}{\epsilon}

\newcommand{\eqn}[1]{Eq.\,(\ref{#1})}
\newcommand{\eqs}[1]{Eqs.\,(\ref{#1})}
\newcommand{\fig}[1]{Fig.\,\ref{#1}}

\newcommand{\sct}[1]{Sect.\,\ref{#1}}

\newcommand{\dd}{{\rm d}}
\newcommand{\dderiv}[2]{\frac{\dd #1}{\dd #2}}

\newcommand{\order}[1]{{\cal O}(#1)}

\newcommand{\qcd}{\abbrev{QCD}}
\newcommand{\sm}{\abbrev{SM}}
\newcommand{\mssm}{\abbrev{MSSM}}
\newcommand{\susy}{\abbrev{SUSY}}

\newcommand{\nlo}{\abbrev{NLO}}
\newcommand{\nnlo}{\abbrev{NNLO}}
\newcommand{\nll}{\abbrev{NLL}}
\newcommand{\nnll}{\abbrev{NNLL}}
\newcommand{\nklo}[1]{\abbrev{N$^{#1}$LO}}
\newcommand{\nkll}[1]{\abbrev{N$^{#1}$LL}}

\newcommand{\msbar}{\ensuremath{\overline{\text{\abbrev{MS}}}}}
\newcommand{\drbar}{\ensuremath{\overline{\text{\abbrev{DR}}}}}
\newcommand{\drbarprime}{\ensuremath{\overline{\text{\abbrev{DR}}}^\prime}}
\newcommand{\mdrbar}{\ensuremath{\overline{\text{\abbrev{MDR}}}}}
\newcommand{\mdrbarprime}{\ensuremath{\overline{\text{\abbrev{MDR}}}^\prime}}
\newcommand{\DlamSM}{\ensuremath{(\Delta\lambda)_{\bar{\alpha}_t^2\asbar^2}}}
\newcommand{\DlamSMEFT}{\ensuremath{(\Delta\lambda_{\code{EFT}})_{\bar{\alpha}_t^2\asbar^2}}}

\newcommand{\DlamMSSM}{\ensuremath{(\Delta\lambda)_{\alpha_t^2\as^2}}}

\newcommand{\DlamMSSMEFT}{\ensuremath{(\Delta\lambda_{\code{EFT}})_{\alpha_t^2\as^2}}}
\newcommand{\DlamMSSMHtm}{\ensuremath{(\Delta\lambda_{\code{H3m}})_{\alpha_t^2\as^2}}}

\newcommand{\mut}{\mu_t}
\newcommand{\mus}{\mu_S}

\newcommand{\vev}{v}

\newcommand{\mstop}[1]{m_{\tilde{t}_#1}}
\newcommand{\mstopav}{m_{\tilde{t}}}
\newcommand{\msquark}{m_{\tilde{q}}}
\newcommand{\msquarkf}[2]{m_{\tilde{#1}_#2}}
\newcommand{\mgluino}{m_{\tilde{g}}}
\newcommand{\mtop}{m_t}
\newcommand{\Mtop}{M_t}
\newcommand{\CP}{\abbrev{CP}}
\newcommand{\unit}[1]{\,\text{#1}}
\newcommand{\MeV}{\unit{MeV}}
\newcommand{\GeV}{\unit{GeV}}
\newcommand{\TeV}{\unit{TeV}}

\newcommand{\HSSUSY}{\code{HSSUSY}}
\newcommand{\SUSYHD}{\code{SusyHD}}
\newcommand{\FlexibleSUSY}{\code{FlexibleSUSY}}
\newcommand{\DlamShiftDRpToMS}{\ensuremath{(\delta\lambda)_{\alpha_t^2a_s^2}}}
\newcommand{\sinb}[1]{s_\beta^{#1}}
\newcommand{\cSM}{\ensuremath{c_\text{\sm}^{(2,0)}}}

\title{The light \CP-even \mssm\ Higgs mass resummed to fourth logarithmic order}
\author{R.V. Harlander, J. Klappert, A.D. Ochoa Franco, and A. Voigt}
\affil{Institute for Theoretical Particle Physics and Cosmology, RWTH
  Aachen University, 52074 Aachen, Germany}
\date{}

\begin{document}
\maketitle
\thispagestyle{empty}
\begin{abstract}
  We present the calculation of the light neutral \CP-even Higgs mass in
  the \mssm\ for a heavy \susy\ spectrum by resumming enhanced terms
  through fourth logarithmic order (\nkll{3}), keeping terms of leading
  order in the top Yukawa coupling $\alpha_t$, and \nnlo\ in the strong
  coupling $\alpha_s$. To this goal, the three-loop matching coefficient
  for the quartic Higgs coupling of the \sm\ to the \mssm\ is
  derived to order $\alpha_t^2\alpha_s^2$ by comparing the perturbative
  \eft\ to the fixed-order expression for the Higgs mass. The new
  matching coefficient is made available through an updated version of
  the program \himalaya. Numerical effects of the higher-order
  resummation are studied using specific examples, and sources of
  theoretical uncertainty on this result are discussed.

\end{abstract}

\begin{textblock*}{10em}(\textwidth,1.5cm)
\raggedright\noindent
\RHheaderline
\end{textblock*}

\clearpage
\tableofcontents
\clearpage

\section{Introduction}

In the \mssm\ (the minimal supersymmetric (\susy) extension of the
Standard Model (\sm)), the mass of the lightest \CP-even Higgs boson is
predicted to be of the order of the electroweak scale.  More precisely,
at the tree-level, the Higgs boson mass is restricted to be smaller than
or equal to the mass of the $Z$ boson, $M_h \leq M_Z$. In viable
parameter regions of the \mssm, the loop corrections to the mass of the
light \CP-even Higgs boson must therefore be large in order for the
\mssm\ to accommodate for the measured Higgs mass value of
\cite{Aad:2015zhl}
\begin{equation}
  M_h = (125.09\pm 0.32) \GeV .
  \label{eq:mhval}
\end{equation}
It has been known for a long time that these loop corrections are indeed
large, predominantly due to contributions from top quarks and their
super-partners, the
``stops''\,\cite{Haber:1990aw,Okada:1990vk,Ellis:1990nz,Ellis:1991zd,
  Barbieri:1991tk,Chankowski:1992er,Dabelstein:1994hb}.  To be specific,
in the limit where the superpartners are much heavier than the
electroweak scale, the pole mass of the light \CP-even Higgs boson,
including the dominant \one-loop contribution, reads
\cite{Allanach:2004rh}
\begin{align}
  \mhiggs^2 = M_Z^2 \cos^2 2 \beta + 
  \frac{6 g_t^4 v^2}{(4\pi)^2} \left[ \ln \frac{\mstopav^2}{m_t^2} +
  \frac{X_t^2}{\mstopav^2} - \frac{X_t^4}{12 \mstopav^4}\right]
  \label{eq:oneLstop},
\end{align}
where $m_t$ is the top-quark mass, $\mstopav^2 = \mstop{1}\mstop{2}$ is
the average of the two stop masses $\mstop{i}$ ($i=1,2$), $g_t$ is the
\sm\ top Yukawa coupling, $v\sim 246\GeV$ is the vacuum expectation
value of the \sm, $X_t = A_t - \mu/\tan\beta$ is the stop mixing
parameter, $A_t$ is the trilinear Higgs--stop coupling, $\mu$ is an
MSSM superpotential parameter and $\tan\beta = v_u/v_d$ is the ratio
of the up- and down-type MSSM Higgs boson \abbrev{VEV}s.
\eqn{eq:oneLstop} illustrates that a heavy \susy\ spectrum
logarithmically enhances the corrections to the Higgs mass, and that the
effect of the stop mixing parameter maximally enhances the Higgs mass at
$|X_t/\mstopav|=\sqrt{6}$.  Including higher order effects, it turns out
that the stop masses must be larger than $\mstop{i} \gtrsim 1\TeV$ in
order to predict the physical Higgs mass of \eqn{eq:mhval} in scenarios
with degenerate \susy\ mass parameters and arbitrary stop
mixing~\cite{Bagnaschi:2014rsa,Vega:2015fna,Bahl:2016brp,Bahl:2017aev,Allanach:2018fif}.

For stop masses larger than about $1\TeV$, logarithmic corrections like
the $\ln(\mstopav^2/m_t^2)$ term in \eqn{eq:oneLstop} may spoil the
precision of the perturbative fixed-order result. However, using an
effective field theory (\eft) approach, the leading (next-to-leading,
etc.\@) powers of these logarithmic terms can be resummed to all orders
in the coupling constants.  Terms of order $v^2/\ms^2$, where $\ms$ is
the typical \susy\ particle mass, are usually neglected in an
\eft\ calculation, which is justified at $\ms\gtrsim
1$\,TeV~\cite{Bahl:2017aev}. Their inclusion can be achieved by taking
into account higher-dimensional operators\,\cite{Bagnaschi:2017xid}, or
through so-called ``hybrid''
approaches~\cite{Hahn:2013ria,Athron:2017fvs,Bahl:2016brp,Bahl:2017aev,Athron:2016fuq,
  Staub:2017jnp,Bahl:2018jom}.

The resummation of the logarithmic terms through an \eft\ calculation is
achieved by integrating out the \susy\ partners at a high scale
$\mus\sim\ms$. This means that the \msbar\ parameters of the effective
theory (the \sm), in particular the quartic Higgs coupling $\lambdabar$,
which itself is \textit{not} a free \mssm\ parameter, are expressed in
terms of the \mssm\ parameters at that scale. The \sm\ parameters are
then evolved down to a low scale $\mut\sim v$ through numerical
\sm\ renormalization group running, which implicitly resums all
logarithms of ratios of the high and the low scale, $\mus/\mut$. This
allows to evaluate the Higgs pole mass within the \sm\ in terms of
\sm\ parameters:
\begin{align}
  \mhiggs^2 = \lambdabar(\mut) \vbar^2(\mut) + \cdots\,,
  \label{eq:Mh_SM}
\end{align}
where $\vbar$ is the vacuum expectation value of the Higgs field in the
\msbar\ scheme, and the ellipsis denotes terms of higher order in the
\sm\ couplings.

The crucial ingredients in the \eft\ approach are therefore the running
\mssm\ parameters, which can be obtained from spectrum generators such
as \FlexibleSUSY\ \cite{Athron:2014yba,Athron:2017fvs},
\code{SARAH}/\code{SPheno}\ \cite{Porod:2003um,Staub:2009bi,
  Porod:2011nf,Staub:2010jh,Staub:2012pb, Staub:2013tta,Staub:2017jnp},
\code{SOFTSUSY}\ \cite{Allanach:2001kg,Allanach:2014nba}, or
\code{SuSpect}\ \cite{Djouadi:2002ze}, the $\beta$ functions of the
\sm\ parameters, and the matching relations of the \sm\ to the
\mssm\ parameters. In order to consistently resum through first
(leading), second (next-to-leading), \ldots, $k^\text{th}$ logarithmic
order (\abbrev{LL}, \nll, \ldots, \nkll{k-1}), one needs to take into
account the $\beta$ function of the quartic Higgs coupling,
$\beta_\lambda$, through $k$-loop order, and the corresponding matching
coefficient $\Delta\lambda$ through $(k-1)$-loop order, while for the
other parameters, the corresponding functions are required only at lower
orders.
While $\beta_\lambda$ is known through \four\
loops~\cite{Martin:2015eia,Chetyrkin:2016ruf}, however, the matching coefficient
$\Delta\lambda$ has been available only through \two\ loops
\cite{Draper:2013oza,Bagnaschi:2014rsa,Vega:2015fna,Bagnaschi:2017xid}.  The
logarithmic order for the resummed expression of the Higgs mass has
thus been limited to the third logarithmic order (\nnll) up to now.

In this paper, we show how the \three-loop matching coefficient for the
quartic Higgs coupling can be extracted from the \three-loop fixed-order
expression \cite{Harlander:2008ju,Kant:2010tf} for the Higgs pole mass
in the \mssm.  The latter has recently been implemented into the
\himalaya\ library \cite{Harlander:2017kuc}.  We make the \three-loop
threshold correction to the quartic Higgs coupling available in
\himalaya~2.0.1, which can be downloaded from
\begin{center}
  \url{https://github.com/Himalaya-Library}
\end{center}
This result allows us to study the impact of the resummation to fourth
logarithmic order on the numerical prediction of the Higgs boson mass in
the decoupling limit of the \mssm\ by implementing the \three-loop
correction into \HSSUSY, an \eft\ spectrum generator from the
\FlexibleSUSY\ package.

\section{Formalism}

As briefly described in the introduction, there are different
approximation schemes commonly used to calculate the light \CP-even
Higgs boson mass in the \mssm: The fixed-order, the \eft, and the hybrid
calculation.  The fixed-order calculation includes the \susy\ effects
through an expansion in terms of couplings up to a fixed order.  In this
expansion, logarithmic corrections appear, which may be large if there
is a large split between the \susy\ and the electroweak scale, $\ms \gg
v$.  The fixed-order calculation is therefore a suitable approximation
as long as $\ms \sim v$.  In an \eft\ calculation, an expansion in
powers of $\vbar^2/\ms^2$ is performed, and the leading (sub-leading,
\ldots) powers of such logarithms are resummed to all orders in the
couplings.  An \eft\ calculation is therefore a suitable approximation
if $\ms \gg v$, but becomes invalid when $\ms \sim v$.

In the following sections, we describe both the fixed-order and the
\eft\ calculation in more detail, in order to prepare for the extraction
of the \three-loop correction to the quartic Higgs coupling of the
Standard Model later in \sct{sec:extraction_of_lambda}.

The set of \sm\ \msbar\ parameters relevant to our calculation will be
denoted as
\begin{equation}
  \bar X = \{\lambdabar, \bar\alpha_t, \asbar, \vbar\} \,,
  \label{eq:SM_parameters}
\end{equation}
where
\begin{align}
  \bar\alpha_t &= \frac{\bar{g}_t^2}{4\pi} , &
  \asbar &= \frac{\bar{g}_3^2}{(4\pi)^2} ,
\end{align}
$\lambdabar$ denotes the quartic Higgs coupling, $\bar{g}_t$ the
\sm\ top Yukawa coupling, $\bar{g}_3$ the strong gauge coupling, and
$\vbar$ the vacuum expectation value of the Higgs field in the \sm.
Furthermore, we use the following set of \mssm\ parameters, renormalized
in the \drbarprime\ scheme~\cite{Jack:1994rk},
\begin{align}
  Y = \{\alpha_t,\as,\vev,\mstop{1},\mstop{2},X_t,\mgluino,\msquark\}\,,
\end{align}
with
\begin{align}
  \alpha_t &= \frac{y_t^2}{4\pi} \,, &
  \as &= \frac{g_3^2}{(4\pi)^2} \,, &
  v &= \sqrt{v_u^2 + v_d^2} \,, &
  \msquark &= \left( \prod_{f\in\{u,d,c,s,b\}} \prod_{n=1}^2 \msquarkf{f}{n} \right)^{1/10} \,,
\end{align}
whereas $y_t$ denotes the \mssm\ top Yukawa coupling, $g_3$ the strong
gauge coupling, $v_u$ and $v_d$ the vacuum expectation values of the
neutral up- and down-type Higgs bosons, $X_t = A_t - \mu/\tan\beta$ the
stop mixing parameter, $\mgluino$ the gluino mass, and $\msquark$ the
average mass of all squarks but the stops.  The running stop masses
$\mstop{1}\leq \mstop{2}$ are the eigenvalues of the stop mass matrix:
\begin{equation}
  \begin{split}
    \mathcal{M}_t = \left(\begin{array}{cc}
      m_t^2 + m_{Q,3}^2 & m_t X_t\\
      m_t X_t & m_t^2 + m_{U,3}^2
    \end{array}\right)\,,
    \label{eq:mQU}
  \end{split}
\end{equation}
with the \susy\ breaking parameters $m_{Q,3}$ and $m_{U,3}$.  Note that,
due to the \susy\ constraints, $Y$ does not contain a separate parameter
for the quartic Higgs coupling.

\subsection{Fixed-order calculation}

In the Standard Model, the pole mass of the Higgs boson can be expressed
as a series expansion in terms of the \sm\ couplings and logarithms.
The dominant terms in the expansion are those which involve the strong
and the top Yukawa coupling.  In the following, we consider only
corrections to the tree-level Higgs mass of the form
$\order{\bar\alpha_t^2 \asbar^n}$ with $n \geq 0$, in which case the
pole mass of the Higgs boson can be expressed in terms of
\msbar\ parameters as
\begin{equation}
  \mhiggs^2 = \vbar^2(\mut)\left[\lambdabar(\mut) +
    \kappa\bar\alpha_t^2(\mut)\sum_{n=0}^\infty\sum_{p=0}^{n+1}
    \kappa^n\asbar^n(\mut)
    c^{(n,p)}_\text{\sm}\,\barlmut^p\right]\,,\\
  \label{eq:mh2sm}
\end{equation}
where
\begin{align}
    \barlmut &= \ln\frac{\mut^2}{\mtbar^2}\,, &
    \mtbar^2 &= \frac{\bar{g}_t^2 \vbar^2}{2} = 2\pi\bar\alpha_t\vbar^2\,,  
\end{align}
and $\mut$ is the renormalization scale.  The auxiliary parameter
$\kappa=1$ has been introduced to label the orders of perturbation
theory. The $c^{(n,p)}_\text{\sm}$ are pure numbers; through \three-loop order ($n=2$),
the non-logarithmic coefficients
read~\cite{Martin:2007pg,Martin:2014cxa,Martin:2015eia}
\begin{equation}
  \begin{split}
    c^{(0,0)}_\text{\sm} &= c^{(1,0)}_\text{\sm} = 0\,,\\
    c^{(2,0)}_\text{\sm} &= 
      -\frac{1888}{9}+160 \zeta_3+\frac{7424}{45} \zeta_2^2
      -\frac{1024}{3}
      \text{Li}_4\left(\frac{1}{2}\right)
      -\frac{512}{9}\text{Li}_2^2\left(\frac{1}{2}\right)
      -\frac{1024}{9} \text{Li}_2\left(\frac{1}{2}\right) \zeta_2 \,,
    \label{eq:csm}
  \end{split}
\end{equation}
where
\begin{equation}
  \begin{split}
    \zeta_2=\frac{\pi^2}{6}&=1.64493\ldots\,,\qquad
    \zeta_3 = 1.20206\ldots\,,\\
    \text{Li}_2\left(\frac{1}{2}\right) &= 0.582241\ldots\,,\qquad
    \text{Li}_4\left(\frac{1}{2}\right) = 0.517479\ldots\,.
    \label{eq:zetali}
  \end{split}
\end{equation}
The logarithmic coefficients ($p\neq 0$) can be easily obtained from the
renormalization-group (\abbrev{RG}) invariance of $\mhiggs^2$ and the
\abbrev{RG}-equations (\abbrev{RGE}s) of the parameters
\cite{Martin:2007pg},
\begin{equation}
  \begin{split}
    \mu\dderiv{}{\mu} \bar x_i(\mu) = \beta_{\bar x_i}(\bar X(\mu))\,,
    \label{eq:rge}
  \end{split}
\end{equation}
with $\bar{x}_i \in \bar{X}$.  The terms in the \sm\ $\beta$ functions
that are relevant for our discussion read
\begin{equation}
  \begin{split}
    \beta_{\asbar} &= -14\kappa\asbar^2 - 52\kappa^2\asbar^3 + \cdots\,, \\
    \beta_{\bar\alpha_t} &= -\bar\alpha_t\left[16\kappa\asbar +
      216(\kappa\asbar)^2 + 1238.7(\kappa\asbar)^3 + \cdots\right]\,,\\
    \beta_{\lambdabar} &=
    -\kappa\bar\alpha_t^2\left[12 + 64\kappa\asbar +
    8\left(\frac{133}{3}-16\zeta_3\right)(\kappa\asbar)^2
    -16616.3 (\kappa\asbar)^3 + \cdots\right]\,.
    \label{eq:betasm}
  \end{split}
\end{equation}

In the \mssm\ one can write an analogous expression for the light
\CP-even Higgs boson mass in terms of the \mssm\ parameters.  Neglecting
sub-leading terms of $v^2/\ms^2$, one obtains the expansion in the
decoupling limit, which reads
\begin{equation}
    \mhiggs^2 = M_Z^2\cos^22\beta +
    \kappa\vev^2(\mut)\alpha_t^2(\mut)\sinb{4}\sum_{n=0}^\infty\sum_{p=0}^{n+1}\kappa^n
    \as^n(\mut)
    c^{(n,p)}_\text{\mssm}(Y(\mut))\,\lmut^p\,,
    \label{eq:mh2mssm}
\end{equation}
with 
\begin{align}
  \lmut &= \ln\frac{\mut^2}{\mtop^2}\,, &
  \mtop^2 &= \frac{y_t^2 v_u^2}{2} = 2\pi\alpha_t v_u^2
  = 2\pi\alpha_t\vev^2\sinb{2}\,, & \sinb{} = \sin\beta.
\end{align}
The coefficients $c^{(n,p)}_\text{\mssm}$ have been calculated
analytically through $n=1$ and can be extracted from
\citeres{Degrassi:2001yf,Martin:2001vx,Martin:2002iu,Martin:2002wn}. The
result for $n=2$ was obtained in \citere{Harlander:2008ju,Kant:2010tf}
in terms of ``hierarchies'', i.e., expansions in various limits of the
\mssm\ particle spectrum.\footnote{As has been shown recently, the
  \three-loop calculation of the Higgs mass in the \mssm\ in the
  \drbarprime\ scheme is consistent with
  supersymmetry~\cite{Capper:1979ns,Stockinger:2005gx,Stockinger:2018oxe};
  see also \citeres{Harlander:2009mn,Harlander:2006xq} concerning the
  consistency of dimensional reduction\,\cite{Siegel:1979wq} and
  perturbative calculations in \susy.} The $c^{(n,p)}_\text{\mssm}$
contain logarithmic terms of the form $\ln( \mtop/\ms)$ which spoil the
convergence properties of the purely fixed-order result of
\eqn{eq:mh2mssm} if $\ms\gg \mtop$.  To make this more explicit, let us
introduce a second scale $\mus\neq \mut$ by perturbatively evolving the
running \mssm\ parameters in \eqn{eq:mh2mssm} from $\mut$ to $\mus$,
using the corresponding $\beta$ functions defined in analogy to
\eqn{eq:rge}.  This means that we apply the replacement
\begin{equation}
  \begin{split}
    y_i(\mut)=y_i(\mus) + \sum_{n=1}^\infty\sum_{p=1}^n\kappa^n
    d_i^{(n,p)}(Y(\mus))\lts^p\,,\qquad
    \lts = \ln\frac{\mut^2}{\mus^2}
    \label{eq:rgesol}
  \end{split}
\end{equation}
to \eqn{eq:mh2mssm} for all \mssm\ parameters $y_i\in Y$, where the
$d_i^{(n,p)}$ are determined by the perturbative coefficients of the
respective $\beta$ functions. After re-expanding in $\kappa$, this
results in a relation of the form
\begin{equation}
    \mhiggs^2 = M_Z^2\cos^22\beta 
     +\kappa\vev^2(\mus)\alpha_t^2(\mus)\sinb{4}\sum_{n=0}^\infty\sum_{p=0}^{n+1}
    \sum_{k=0}^{n+1-p}\kappa^n\as^n(\mus)
    c^{(n,p,k)}_\text{\mssm}(Y(\mus))\,\lmut^p\lts^k\,.
    \label{eq:mhfomus}
\end{equation}
In a fixed-order calculation, the perturbative expansion is truncated at
finite order in $\kappa$.  Keeping terms through order $\kappa^N$, we
will denote this result as
\begin{equation}
  \begin{split}
    M_{h,\text{\fo},N}^2(\mut,\mus)\,.
    \label{eq:mhfon}
  \end{split}
\end{equation}
For $\mtop\ll \ms$, any choice of $\mut$ and $\mus$ will result in large
logarithms in \eqn{eq:mhfon}. This is avoided in the \eft\ approach
which allows to resum the (leading, sub-leading, etc.\ powers of)
logarithms $\lts$ to all orders in perturbation theory. This will be the
subject of the next section. Of course, a re-expansion of the
\eft\ result must take the fixed-order form of \eqn{eq:mhfon}
again. Comparison of this re-expanded result to the fixed-order
\three-loop result will allow us to derive the \three-loop matching
coefficient for $\bar\lambda$ in \sct{sec:extraction_of_lambda}.

\subsection{EFT calculation}\label{sec:eft_calculation}

The idea behind the \eft\ calculation is to resum the logarithms of the
form $\lts$ in \eqn{eq:mhfomus} (``large logarithms'') by integrating
out the heavy (i.e., \susy) particles.  As a result, one obtains a
relation between the parameters of the effective theory (the \sm) and
the full theory (the \mssm) of the form
\begin{equation}
  \bar x_i(\mu) = f_i(Y(\mu),\mu) \,.
  \label{eq:matchgeneric}
\end{equation}
In particular, one obtains a relation between $\lambdabar$ and the
\mssm\ parameters, which means that the Higgs mass in the \sm, given by
\eqn{eq:mh2sm}, is fixed in terms of the parameters $Y$.  The $f_i$ in
\eqn{eq:matchgeneric} are known in terms of perturbative expansions,
neglecting terms of the order $v^2/\ms^2$.  They depend
explicitly on the renormalization scale $\mu$ in the form of
$\ln(\mu/\ms)$.  Therefore, if \eqn{eq:matchgeneric} is employed at the
scale $\mu \sim \ms$, no large logarithms appear in the matching.  For
our purpose, the relevant matching relations of \eqn{eq:matchgeneric}
take the form
\begin{equation}
  \begin{split}
    \lambdabar &= \frac{M_Z^2}{\vev^2}\cos^22\beta +
    \kappa\alpha_t^2\sinb{4}(\Delta\lambda)_{\alpha_t^2}
      +\kappa^2\alpha_t^2\as\sinb{4}(\Delta\lambda)_{\alpha_t^2\as}
      +\kappa^3\alpha_t^2\as^2\sinb{4}\DlamMSSM + \cdots\,,
\\
    \asbar &= \as\left(1
    + \kappa\as(\Delta \as)_{\as}
    + \kappa^2\as^2(\Delta \as)_{\as^2} + \cdots\right)\,,
\\
    \bar\alpha_t &= \alpha_t\sinb{2}\left(1
      + \kappa\as(\Delta \alpha_t)_{\as}
      + \kappa^2\as^2(\Delta \alpha_t)_{\as^2} + \cdots\right)\,,
\\
    \vbar &= \vev + \cdots\,,
  \end{split}%
  \label{eq:match}%
\end{equation}
where the perturbative coefficients $(\Delta x_i)$ can be found in
\citeres{Bagnaschi:2014rsa,Bednyakov:2002sf,Bednyakov:2005kt}, except
for $\DlamMSSM$, which will be one of the central results of this
paper. Explicit expressions for the degenerate-mass case will be given
in \sct{sec:degmass}. The dependence on the renormalization scale $\mu$,
indicated in \eqn{eq:matchgeneric}, has been suppressed here.

Assuming that the numerical values for the $y_i(\mus\sim\ms)$ are
known,\footnote{In practice, they are obtained from a spectrum
  generator, using a specific \mssm\ scenario, constrained by the
  experimental values for the \sm\ parameters; see also
  \sct{sec:numerics}.}  \eqn{eq:matchgeneric} provides numerical values
for the \msbar\ \sm\ parameters $\bar x_i(\mus)$. Then one may use the
numerical solution of the \sm\ \msbar\ \abbrev{RGEs} of \eqn{eq:rge} to
evolve the $\bar x_i(\mus)$ down to $\mut\sim\Mtop$.  In solving the
\abbrev{RGEs} numerically, one effectively resums large logarithms of
the form $\lts=\ln(\mut/\mus)$.  This is in contrast to the fixed-order
calculation, where these large logarithms appear explicitly in
$\mhiggs^2$ up to a fixed order, see \eqn{eq:mhfon}.  The $\bar
x_i(\mut)$ are then inserted into \eqn{eq:mh2sm} in order to calculate
$M_h^2$ up to terms of order $v^2/\ms^2$.
We denote this result as
\begin{equation}
  \begin{split}
    M_{h,\text{\eft}}^2(\mut,\mus)\,.
  \end{split}
\end{equation}
The only fixed-order logarithms involved in this result are of the
form $\ln(\mus/\ms)$ from \eqn{eq:matchgeneric}, and
$\ln(\mut/\mtbar)$ from \eqn{eq:mh2sm}. They can be made small by
choosing $\mus\sim\ms$ and $\mut\sim\mtbar$, respectively.

\subsection{Re-expanding the \eft\ result}

The perturbative version of the approach described in the previous
section would be to first evolve the $\bar x_i(\mu)$ perturbatively from
$\mu=\mut$ to $\mus$, i.e., to solve \eqn{eq:rge} in the form
\eqn{eq:rgesol}, which explicitly introduces large logarithms of the
form $\lts$:
\begin{equation}
  \begin{split}
    \mhiggs^2 &= \vbar^2(\mus)\left[\lambdabar(\mus) +
    \kappa\bar\alpha_t^2(\mus)\sum_{n=0}^\infty\sum_{p=0}^{n+1}
    \sum_{k=0}^{n+1-p}\kappa^n\asbar^n(\mus)
    c^{(n,p,k)}_\text{\sm}\,\lmut^p\lts^k\right]\,.
  \end{split}
  \label{eq:Mh2_eft_cut_off}
\end{equation}
Subsequently, one expresses the $\bar x_i(\mus)$ by the $y_i(\mus)$
through \eqn{eq:matchgeneric}. This last step only introduces small
logarithms of the form $\ln(\mus/\ms)$. Re-expanding in $\kappa$, one
thus arrives at a result which coincides with \eqn{eq:mhfomus}.
If we keep terms through order $\kappa^N$, this result will be denoted
as
\begin{equation}
  \begin{split}
    M^2_{h,\text{\eft},N}(\mut,\mus)\,.
    \label{eq:mheftn}
  \end{split}
\end{equation}
Obviously, the following formal relation applies:
\begin{equation}
  \begin{split}
    M_{h,\text{\eft}}^2(\mut,\mus) =
    M_{h,\text{\eft},N}^2(\mut,\mus) + \order{\kappa^{N+1}}\,,
  \end{split}
\end{equation}
if the same order in the perturbative expansions of the
$\beta$-functions, the matching relations, and the \sm\ expression for
$M_h^2$ is used in deriving the results on both sides of this
equation. Since the perturbative expression for $M_h^2$ is unique, we
also have
\begin{equation}
  \begin{split}
    M_{h,\text{\fo},N}^2(\mut,\mus)
    = M_{h,\text{\eft},N}^2(\mut,\mus) \,,
    \label{eq:foeqeft}
  \end{split}
\end{equation}
with the fixed-order result of \eqn{eq:mhfon}.  These relations will be
used in the next section to extract the three-loop matching relation for
the quartic Higgs coupling $\lambdabar(\mus)$.

The goal of this paper is to calculate the light \CP-even Higgs pole
mass of the \mssm\ in the decoupling limit including the fixed-order
through $\order{\alpha_t^2\as^2}$ (\nklo{3}), as well as resummation in
$\alpha_t^2\alpha_s^n$ through fourth logarithmic order (\nkll{3}).
This calculation requires to include
\begin{itemize}
\item the \four-loop $\beta$ function for $\lambdabar$ to order $\kappa^4\bar{\alpha}_t^2\asbar^3$;
\item the \three-loop $\beta$ function for $\bar{\alpha}_t$ to order $\kappa^3\bar{\alpha}_t\asbar^3$;
\item the \two-loop $\beta$ function for $\asbar$ to order $\kappa^2\asbar^3$;
\item the \three-loop matching relation for $\lambdabar$ to order $\kappa^3\bar{\alpha}_t^2\asbar^2$;
\item the \two-loop matching relation for $\bar{\alpha}_t$ to order $\kappa^2\bar{\alpha}_t\asbar^2$;
\item the \one-loop matching relation for $\asbar$ to order $\kappa\asbar^2$;
\item the \three-loop \sm\ contributions to the Higgs mass,
  \eqn{eq:mh2sm}, to order $\kappa^3\bar{\alpha}_t^2\asbar^2$.
\end{itemize}
Currently, all of the necessary expressions are known, except for the
\three-loop matching relation for $\lambdabar$ to order
$\bar{\alpha}_t^2\asbar^2$.  In the next section, we will derive this
quantity from the \htm\ result, i.e., the known fixed-order corrections
of $\order{\alpha_t^2\as^2}$ for $M_h^2$ from
\citeres{Harlander:2008ju,Kant:2010tf}.

\section{Extraction of the three-loop matching coefficient}
\label{sec:extraction_of_lambda}

\subsection{General procedure}

Using \eqs{eq:mh2sm}, \noeqn{eq:csm}, \noeqn{eq:betasm} and
\noeqn{eq:match}, and setting $\mut=\mus$, the \three-loop \susy\ \qcd\ result for
$M_{h,\text{\eft},3}^2(\mus,\mus)$ can be written in the following form:
\begin{equation}
  \begin{split}
	M_{h,\text{\eft},3}^2&(\mus,\mus) = 
	M_{h,\text{\eft},2}^2(\mus,\mus)\\&
        +\kappa^3\vev^2\alpha_t^2a_s^2\sinb{4}\Bigg\{
	   368\;\lst^3 + \Big[80 + 48(\Delta a_s)_{a_s} 
	   + 96(\Delta\alpha_t)_{a_s}\Big]\lst^2\\&\qquad
	- \Big[64\zeta_3 + \frac{1028}{3} + 16(\Delta a_s)_{a_s} + 128(\Delta \alpha_t)_{a_s}\\&\qquad\qquad
		- 6(\Delta\alpha_t)_{a_s}^2
                - 12 (\Delta\alpha_t)_{a_s^2}\Big] \lst
        \\&\qquad
      + 16(\Delta \alpha_t)_{a_s} 
		- 9(\Delta \alpha_t)_{a_s}^2
		- 6 (\Delta\alpha_t)_{a_s^2}
	+ \DlamMSSM + \cSM\Bigg\},
	\label{eq:mh2_3L_eft}
  \end{split}
\end{equation}
where $\lst=\ln(\mus^2/\mtop^2)$ and, as before, the $\mus$ dependence of
$\alpha_t$, $a_s$, $\Delta\alpha_t$, $\Delta a_s$ and $\Delta\lambda$
is suppressed.  The only unknown term on the r.h.s.\ of
\eqn{eq:mh2_3L_eft} is the \three-loop matching coefficient for the
quartic Higgs coupling $(\Delta\lambda)_{\alpha_t^2a_s^2}$. Assuming
that the \three-loop fixed-order result
$M_{h,\text{\fo},3}^2(\mus,\mus)$ is known, we could insert
\eqn{eq:foeqeft} into \noeqn{eq:mh2_3L_eft} and solve for the unknown
matching coefficient:
\begin{equation}
  M_{h,\text{\fo},3}^2(\mus,\mus) - 
  M_{h,\text{\eft},3}^2(\mus,\mus)\bigg|_{(\Delta\lambda)_{\alpha_t^2a_s^2}=0}
  = \kappa^3\vev^2\alpha_t^2\as^2\sinb{4}(\Delta\lambda)_{\alpha_t^2a_s^2}\,.
  \label{eq:deltalambda}
\end{equation}
Note that all large logarithms $\lst$ cancel on the l.h.s.\ of
\eqn{eq:deltalambda}. Thus, we may write \eqn{eq:deltalambda} as
\begin{equation}
  \begin{split}
    \kappa^3\vev^2\alpha_t^2\as^2\sinb{4}(\Delta\lambda)_{\alpha_t^2a_s^2} = 
    M_{h,\text{\fo},3}^2(\mus,\mus) - 
    M_{h,\text{\eft},2}^2(\mus,\mus)
    - \Delta M_{h,3}^2(\mus)\,,
 \end{split}
  \label{eq:mh2_eft_subtraction}
\end{equation}
where
\begin{equation}
  \begin{split}
    \Delta M_{h,3}^2(\mus) = 
    \kappa^3\vev^2\alpha_t^2\as^2\sinb{4}\Bigg[16(\Delta
      \alpha_t)_{a_s} - 9(\Delta \alpha_t)_{a_s}^2
      -6(\Delta\alpha_t)_{a_s^2} + \cSM
      \Bigg]\,.
    \label{eq:subtract}
  \end{split}
\end{equation}
The matching coefficient $\DlamMSSM$ obtained in this way is defined
in the \msbar\ scheme and expressed in terms of the \mssm\
\drbarprime\ parameters $\alpha_t$ and $\as$, in accordance with
\eqn{eq:match}.\footnote{To convert $\DlamMSSM$ from the \msbar\ to
  the \drbarprime\ scheme, an additional explicit \three-loop
  conversion term of $\order{\alpha_t^2\as^2}$ for $\lambda$ would be
  necessary, analogous to the \one-loop conversion terms of
  \citeres{Martin:1993yx,Summ:2018oko}.}  Inverting the matching
relations for $\alpha_t$ and $\as$,
\begin{equation}
  \begin{split}
    \as &= \asbar\left\{1 - \kappa\asbar\left[\Delta\as)_{\as} -
      \kappa^2\asbar^2((\Delta\as)_{\as^2} -
      2(\Delta\as)_{\as}\right]\right\}\,,\\
    \alpha_t\sinb{2} &= \bar\alpha_t\left\{1 -
    \kappa\asbar(\Delta\alpha_t)_{\as} -
      \kappa^2\asbar^2\left[(\Delta\alpha_t)_{\as^2} -
      (\Delta\as)_{\as}(\Delta\alpha_t)_{\as}
        - (\Delta\alpha_t)^2_{\as}\right]\right\}\,,
  \end{split}
\end{equation}
it can also be expressed in terms of \sm\ \msbar\ parameters according
to
\begin{align}
   \lambdabar &= \frac{M_Z^2}{\vbar^2}\cos^22\beta +
    \kappa\bar\alpha_t^2(\Delta\lambda)_{\bar\alpha_t^2}
      +\kappa^2\bar\alpha_t^2\bar{a}_s(\Delta\lambda)_{\bar\alpha_t^2\bar{a}_s}
      +\kappa^3\bar\alpha_t^2\bar{a}_s^2 \DlamSM
     + \cdots\,,
\end{align}
where
\begin{equation}
  \begin{split}
  (\Delta\lambda)_{\bar\alpha_t^2} &=
  (\Delta\lambda)_{\alpha_t^2}\,,\\
  (\Delta\lambda)_{\bar\alpha_t^2\bar{a}_s} &=
    (\Delta\lambda)_{\alpha_t^2a_s} - 2(\Delta\lambda)_{\alpha_t^2}
    (\Delta\alpha_t)_{a_s}\,,\\
  \DlamSM &= \DlamMSSM + \DlamShiftDRpToMS \,,
  \end{split}
\end{equation}
and
\begin{align}
  \begin{split}
    \DlamShiftDRpToMS &=
    -(\Delta\lambda)_{\alpha_t^2\as}\left[(\Delta a_s)_{a_s} +
      2(\Delta \alpha_t)_{a_s}\right]\\&\qquad
    + (\Delta\lambda)_{\alpha_t^2}\left[
            3(\Delta\alpha_t)^2_{a_s}
            -2(\Delta \alpha_t)_{a_s^2}
            +2(\Delta\alpha_t)_{\as}(\Delta\as)_{\as}\right]\,.
   \end{split}
    \label{eq:shift_DRbarprime_to_MSbar}
\end{align}

\subsection{Three-loop fixed-order result}\label{sec:3fo}

\eqn{eq:deltalambda} shows how the \three-loop matching coefficient for
the quartic Higgs coupling can be extracted from the \three-loop
fixed-order result for the \mssm\ Higgs mass. The latter has been
calculated in \citeres{Harlander:2008ju,Kant:2010tf} in the form of a
set of expansions around various limiting cases for the \susy\ masses
(``hierarchies'').  Since the explicit formul\ae\ for this result are
available in the \code{Mathematica} package \htm\,\cite{h3m}, we will
refer to it as the ``\htm\ result'' in what follows. In all of the
different expansions, terms of $\order{\vev^2/\ms^2}$ have been
neglected.  The calculation was performed in the \drbar\ scheme with an
on-shell renormalization condition for the $\epsilon$-scalars were
$\mep^2=0$.\footnote{The authors also provide their result in a modified
  \drbar\ (\mdrbar) scheme, where heavy \susy\ particles automatically
  decouple.} We refer to this renormalization scheme as the
``\htm\ scheme''.

\subsubsection{Transformation to \drbarprime}

In order to be able to seamlessly combine the \three-loop result in the
\htm\ scheme with existing lower-order calculations, it is necessary to
convert it to the more commonly used \drbarprime\ scheme, where $\mep$
completely decouples from the model.  To do that, we need to reconstruct
the $\mep$-terms in the \htm\ result. This can be done by noting that,
up to \two-loop $\order{\alpha_t^2\as}$, the analytic form of the
corrections to the Higgs mass are identical in the \drbar, the
\drbarprime, and the \htm\ scheme for $\mep=0$. Since the
\drbarprime\ result is independent of $\mep$ to all orders in
perturbation theory, we can convert the known \two-loop
$\order{\alpha_t^2\as}$ \drbarprime\ expression to the \drbar\ scheme by
shifting the stop masses according to
\citeres{Jack:1994rk,Martin:2001vx,Hermann:2011ha}.  Expanding the
resulting expression to $\order{\alpha_t^2\as^2}$ generates all
$\mep$-dependent terms up this order in the \drbar\ scheme. From there,
we can convert the stop masses and $\mep$ to the \htm\ scheme, using the
formul\ae\ of \citere{Kant:2010tf}. This generates a non-vanishing term at
$\order{\alpha_t^2\as^2}$, which is non-zero even when the on-shell
condition $\mep=0$ is applied. For $\mep=0$, this term
reads\footnote{Note that the limit $\mstop{1}\to\mstop{2}$ in
  \eqn{eq:h3mtodrbarprime} is well-defined.}
\begin{equation}
  \begin{split}
    (\Delta \mhiggs^2)_{\htm\to\drbarprime} &= \frac{8\kappa^3\vev^2\alpha_t^2
      a_s^2\sinb{4}}{\mstop{1}^2\mstop{2}^2\Delta_{12}^3} \left[-6\left(1 +
      l_{S{\tilde g}}\right) \mgluino^2 + 10\left(1+l_{S{\tilde
          q}}\right)\msquark^2 + \sum_{i=1}^2 (1+l_{S{\tilde
          t}_i})\mstop{i}^2\right]\\ &\quad \times
    \left[(\Delta_{12}^3 + \Delta_{12}X_t^4)\sum_{i=1}^2\mstop{i}^2
      - 2\Delta_{12}^3X_t^2 + 4 \mstop{1}^2 \mstop{2}^2
      X_t^4\ln\left(\frac{\mstop{2}}{\mstop{1}}\right)\right] \,,
    \label{eq:h3mtodrbarprime}
  \end{split}
\end{equation}
with $l_{Sx} = \ln\left(\mu_S^2/m_x^2\right)$ and $\Delta_{12} =
\mstop{1}^2 - \mstop{2}^2$.  Adding these terms to the \htm\ result
provides the \three-loop Higgs mass corrections in the
\drbarprime\ scheme:
\begin{equation}
  \begin{split}
    \mhiggs^2\Big|_\text{\drbarprime} = 
    \mhiggs^2\Big|_{\htm} +    (\Delta \mhiggs^2)_{\htm\to\drbarprime}\,.
    \label{eq:htmdr}
  \end{split}
\end{equation}
We checked that the resulting \drbarprime\ expression is renormalization
scale independent by using the corresponding stop mass $\beta$ functions
in the \drbarprime\ scheme. Furthermore, we explicitly verified the
cancellation of the $\lst$ terms in \eqn{eq:deltalambda} up to higher
orders in the hierarchy expansions of the \htm\ result.

\subsubsection{Reconstruction of the logarithmic terms}
\label{sec:combining_log_and_nonlog}

After transforming the \htm\ result into the \drbarprime\ scheme
according to \eqn{eq:htmdr}, it can be inserted into
\eqn{eq:deltalambda}. This results in the \three-loop matching
coefficient for the quartic Higgs coupling, expressed in terms of the
\htm-hierarchies defined in \citere{Kant:2010tf}. We denote this result
as $\DlamMSSMHtm$ in what follows.

Due to renormalization group invariance of the \mssm\ Higgs mass, we can
actually derive the logarithmic terms of the form $\ln(\mu^2/\ms^2)$ in
$\Delta\lambda$ for general \mssm\ particle masses by requiring that
\begin{align}
  \mu\dderiv{}{\mu}
  \left[
    M_{h,\text{\fo},2}^2(\mu,\mu) +
    \Delta M_{h,3}^2(\mu,\mu) + \kappa^3\vev^2\alpha_t^2(\mu)\as^2(\mu)\sinb{4}
    (\Delta\lambda(\mu))_{\alpha_t^2a_s^2}
  \right] = \mathcal{O}(\kappa^4)\,,
  \label{eq:reconstruct_logs}
\end{align}
with $\Delta M_{h,3}^2$ from \eqn{eq:subtract}, and using the
\three-loop \mssm\ $\beta$ functions. We refer to the corresponding
matching coefficient which includes the exact mass dependence of the
logarithmic terms reconstructed in this way as $\DlamMSSMEFT$. Note that
only the non-logarithmic term of the fixed-order three-loop result of
\citere{Kant:2010tf} enters this result.  Of course, expanding
$\DlamMSSMEFT$ in terms of the \htm\ hierarchies up to the appropriate
orders, we recover $\DlamMSSMHtm$ as defined above.

\subsection{Example: degenerate-mass case}\label{sec:degmass}

In this paper, we refer to the limit
$m_{U,3}=m_{Q,3}=\mgluino=\msquark=M_S$ as the ``degenerate-mass case'',
where $m_{Q,3}$ and $m_{U,3}$ are soft-breaking parameters of the
Lagrangian introduced in \eqn{eq:mQU}.  Since we have made the $x_t$
dependence explicit in our result and we neglect all but the leading
terms in $\alpha_t \propto m_t^2$, we can set
$\mstop{1} = \mstop{2} = M_S$ in our expressions.

In the degenerate-mass limit, the expression for
$(\Delta\lambda)_{\alpha_t^2\as^2}$ is simple enough to be quoted
here. In this case, the matching coefficients for the top Yukawa
coupling, defined by \eqn{eq:match}, are given by
\begin{align}
	(\Delta \alpha_t)_{a_s} & = -\frac{8}{3}\left(-1 
		+ L_S + x_t\right) \,, \\
		\begin{split}
	(\Delta\alpha_t)_{a_s^2} & = \frac{2147 - 1844 L_S + 420 L_S^2}{27} + \frac{-928 + 160 L_S}{27}x_t + \frac{16}{9}x_t^2\,,
		\end{split}
	\label{eq:thresholds_degen}
\end{align}
where $L_S = \ln(\mus^2/\ms^2)$. This leads to a subtraction term (see
\eqn{eq:subtract})
\begin{equation}
\begin{split}
  \Delta M_{h,3}^2(\mus) = \kappa^3\vev^2\alpha_t^2a_s^2\sinb{4}
  \Bigg[&-\frac{2\left(2243 - 2228 L_S + 708L_S^2\right)}{9} \\ &-
    \frac{2\left(-1312 + 736L_S\right)x_t}{9} -\frac{224x_t^2}{3} +
    \cSM \Bigg] \,,
  \end{split}
  \label{eq:mh2_3L_eft_degen}
\end{equation}
with $\cSM$ from \eqn{eq:csm}.  Using the
``\texttt{h3} hierarchy'' of \htm, where all \susy\ masses are assumed to be of
comparable size and the expansion is performed in the mass differences,
the \htm\ result for the degenerate-mass case reads
\begin{equation}
\begin{split}
 M_{h,\text{\fo},3}^2 = \frac{8}{27}\kappa^3\vev^2\alpha_t^2a_s^2\sinb{4} \Big[
   &-1246 - 2132 L_S + 1326 L_S^2 - 504 L_S^3 - 1926
   \zeta_3+216L_S\zeta_3 \\ &+ x_t\left(-2776 + 400 L_S - 1464 L_S^2 +
   1908\zeta_3\right)\\ &+ x_t^2\left(3678 - 6L_S + 126
   L_S^2-1485\zeta_3\right) \\ &+ x_t^3\left(2722 +
   20L_S+108L_S^2-2259\zeta_3\right)\Big]+\order{x_t^4}\,,
 \end{split}
 \label{eq:mh2_3L_h3m_degen}
\end{equation}
where we set $\mut=\mus$. Note that higher orders in $x_t$ are not
included in the \htm\ result. The corresponding shift from the \htm\ to
the \drbarprime\ scheme is (see \eqn{eq:h3mtodrbarprime})
\begin{equation}
\begin{split}
    (\Delta \mhiggs^2)_{\htm\to\drbarprime} = 16 \kappa^3\vev^2\alpha_t^2
  a_s^2\sinb{4} \, (1+L_S)\left( 6 - 6 x_t^2 + x_t^4\right) \,.
    \label{eq:h3mtodrbarprime_degen}
\end{split}
\end{equation}
Combining \eqs{eq:mh2_3L_eft_degen}, \noeqn{eq:mh2_3L_h3m_degen}, and
\noeqn{eq:h3mtodrbarprime_degen} according to \eqn{eq:deltalambda}, we
obtain for the matching coefficient in terms of \drbarprime\
parameters
\begin{equation}
  \begin{split}
    (\Delta\lambda(\mus))_{\alpha_t^2\as^2} = \frac{1}{27}\Big\{&6082 -
    27832L_S + 14856L_S^2 -4032L_S^3 \\ &-15408\zeta_3+1728L_S\zeta_3-
    27\cSM \\&+ x_t\left[ 7616L_S -
      11712L_S^2+32(-940+477\zeta_3)\right] \\ &+ x_t^2\left[28848-2640
      L_S + 1008 L_S^2 - 11880\zeta_3 \right] \\ &+ x_t^3\left[160 L_S +
      864 L_S^2 + 8(2722-2259\zeta_3)\right]\Big\}+ \order{x_t^4}\,.
    \label{eq:zeta2}
  \end{split}
\end{equation}
If one re-expresses the \one- and \two-loop corrections in terms of
\sm\ \msbar\ parameters the following shift must be added to
\eqn{eq:zeta2} in the degenerate-mass case,
\begin{equation}
\begin{split}
 (\delta\lambda(\mus))_{\alpha_t^2\as^2} =
	 \frac{1}{27}\Big[&26916L_S - 18816L_S^2-5904L_S^3\\
	 &- x_t\left(-3744+14016L_S+18816L_S^2\right)\\
	 &- x_t^2\left(29652-5424L_S-9936L_S^2\right)\\
	 &-x_t^3\left(-6768 - 13152L_S-2688L_S^2\right)\Big] +\order{x_t^4}\,.
\end{split}
\label{eq:dr_to_ms_shift_degen}
\end{equation}

\subsection{Implementation into \himalaya}

Recently, the original \code{Mathematica}\,\cite{Mathematica}
implementation \htm\ of the \three-loop fixed-order results of
\citere{Kant:2010tf} was translated into the \code{C++} library
\himalaya~1.0~\cite{Harlander:2017kuc} in order to facilitate the
combination of these terms with lower-order codes such as \FlexibleSUSY,
\code{SARAH}/\code{SPheno}, \code{SOFTSUSY} or \code{SuSpect}, which
typically work in the \drbarprime\ scheme.  \himalaya\ 2.0.1 extends the
functionality of \himalaya~1.0 to provide the \three-loop matching
coefficient $\DlamMSSM$ by implementing \eqn{eq:deltalambda}, including
the conversion from the \htm\ to the \drbarprime\ scheme.  In addition,
we implemented the shift of \eqn{eq:shift_DRbarprime_to_MSbar} which
converts the parameters in the matching coefficient from the \drbarprime\ to the
\msbar\ scheme.  This allows to directly use the result in existing
\eft\ codes such as \HSSUSY\,\cite{Athron:2017fvs} or
\SUSYHD\,\cite{Vega:2015fna}, where the \one- and \two-loop corrections
are expressed in terms of \sm\ \msbar\ parameters.

Since the \htm\ result is given as an expansion in mass hierarchies, it
is important to provide uncertainty estimates due to missing higher
order terms in these expansions.  We employ two largely complementary
ways to estimate this uncertainty, referring to the logarithmic and the
non-logarithmic terms, respectively.

Concerning the logarithmic terms, we proceed as follows.  As described
in \sct{sec:combining_log_and_nonlog}, within the \drbarprime\ scheme,
there are two possible extractions of the matching relation for the
quartic Higgs coupling. Both of them use the hierarchy expansions of
\htm\ for the non-logarithmic terms. However, while $\DlamMSSMHtm$ uses
these expansions also for the logarithmic terms, $\DlamMSSMEFT$ contains
their exact mass dependence, derived from \abbrev{RG} invariance (see
\sct{sec:combining_log_and_nonlog}). We thus use the difference of
$\DlamMSSMEFT$ to $\DlamMSSMHtm$ at the scale $\mus$ as an uncertainty estimate:
\begin{equation}
  \delta_\text{exp} = \alpha_t^2\as^2\sinb{4}\left|\DlamMSSMHtm
     - \DlamMSSMEFT\right|.
  \label{eq:log_expansion_uncertainty}
\end{equation}
For the non-logarithmic terms, on the other hand, we consider the
conversion term $\DlamShiftDRpToMS$ defined in
\eqn{eq:shift_DRbarprime_to_MSbar}, whose mass dependence is known
exactly. Since the main source of uncertainty in these expansions occurs
for large mixing, we determine the highest power $n_\text{max}$ of $x_t$
taken into account in the specific \htm\ hierarchy, and use the size of
the terms of order $x_t^n$ with $n_\text{max} < n \leq 4$ in the
non-logarithmic part of
$\DlamShiftDRpToMS$ as uncertainty estimate, named $\delta_{x_t}$.
Note that powers higher than $x_t^{4}$ cannot appear in \DlamMSSM\ when
the result is expressed in terms of the \mssm\ top Yukawa
coupling.  The reason is that the \one-loop correction
$(\Delta\lambda)_{\alpha_t^2}$ contains no terms with
$x_t^{n>4}$, and additional loops involving only (s)quarks, gluons, and
gluinos do
not introduce any additional $X_t$-dependence.
To be specific, let us again consider the limit of degenerate \mssm\ mass
parameters.  In this case, \htm\ uses the \code{h3} hierarchy described
in \sct{sec:degmass}, which includes only terms through $x_t^3$
though. The uncertainty is thus estimated with the help of the
non-logarithmic terms of order $x_t^4$ in
$\DlamShiftDRpToMS$, given by
\begin{equation}
  \begin{split}
  \delta_{x_t} = \frac{1}{27}\alpha_t^2\as^2\sinb{4} \times 5735\,x_t^4 \,.
  \label{eq:xt_uncertainty_deg}
  \end{split}
\end{equation}  
We combine these two uncertainties linearly and define the total
uncertainty due to the hierarchy expansions as
\begin{equation}
  \delta\left(\alpha_t^2\as^2\sinb{4}\DlamMSSMEFT\right) = \delta_{x_t}
     + \delta_\text{exp}.
  \label{eq:delta_lambda_uncertainty_eft}
\end{equation}
Technical details on how to calculate the \three-loop corrections and
the combined uncertainties with \himalaya\ 2.0.1 can be found in
Appendix~\ref{app:himalaya}.

\section{Numerical study and comparison with other calculations}\label{sec:numerics}

To study the numerical impact of the \three-loop matching coefficient
$\DlamSM$ on the value of the light \mssm\ Higgs mass, we have
implemented the coefficient into \HSSUSY, a spectrum generator from the
\FlexibleSUSY\ package which follows the \eft\ approach outlined in
\sct{sec:eft_calculation}. It assumes a high-scale \mssm\ scenario,
where the quartic Higgs coupling of the \sm\ is evaluated at the
\susy\ scale $\mus$ by the matching to the \mssm.  The scenario assumes
that all \susy\ particles have masses around $\ms$ and the Standard
Model is the appropriate \eft\ below that scale.  In the original
version of \HSSUSY, the quartic Higgs coupling is determined using the
\two-loop expressions of
$\order{\bar\alpha_s(\bar\alpha_t+\bar\alpha_b)^2 + (\bar\alpha_t +
  \bar\alpha_b)^3 + \bar\alpha_\tau(\bar\alpha_b + \bar\alpha_\tau)^2}$
from \citeres{Bagnaschi:2014rsa,Bagnaschi:2017xid}, thereby ignoring
terms of $\order{v^2/\ms^2}$.  The known \three- and \four-loop
\sm\ \msbar\ $\beta$ functions of
\citeres{Bednyakov:2012en,Bednyakov:2013eba,Mihaila:2012pz,Buttazzo:2013uya,Bednyakov:2015ooa,Martin:2015eia,Chetyrkin:2016ruf}
are used to evolve the \sm\ parameters to the electroweak scale, where
the gauge and Yukawa couplings as well as the Higgs \abbrev{VEV} are
extracted from the known low-energy observables at full \one-loop level
plus the known \two- and \three-loop \abbrev{QCD} corrections of
\citeres{Fanchiotti:1992tu,Chetyrkin:1999qi,Melnikov:2000qh,Chetyrkin:2000yt}.
The Higgs boson pole mass is calculated by default at the scale $\mut = \Mtop$ at the full
\one-loop level with additional \two-, \three- and \four-loop
\sm\ corrections of $\order{\bar\alpha_s(\bar\alpha_t^2 +
  \bar\alpha_b^2) + (\bar\alpha_t + \bar\alpha_b)^3 +
  \bar\alpha_\tau^3}$, $\order{\bar\alpha_t^4 +
  \bar\alpha_t^3\bar\alpha_s + \bar\alpha_t^2\bar\alpha_s^2}$ and
$\order{\bar\alpha_t^2\bar\alpha_s^3}$ from
\citeres{Degrassi:2012ry,Martin:2014cxa,Martin:2015eia}.\footnote{We thank Thomas Kwasnitza for making
the mixed \two-loop corrections of $\order{\bar\alpha_t^n\bar\alpha_b^m}$ available through \FlexibleSUSY.}
Thus, by
including $\DlamSM$ in the calculation, \HSSUSY\ provides a resummed
Higgs mass prediction in the decoupling limit of the \mssm\ through
\nklo{3}+\nkll{3} at $\order{\bar\alpha_t^2\bar\alpha_s^2}$, including
the full \nlo+\nll\ and the \nnlo+\nnll\ result at
$\order{\bar\alpha_s(\bar\alpha_t^2 + \bar\alpha_b^2) +
  (\bar\alpha_t+\bar\alpha_b)^3 + \bar\alpha_\tau^3}$.
Unless stated otherwise, we set $\mus=\ms$ and $\mut=\Mtop$ in the
following numerical analysis and use $\Mtop = 173.34\GeV$ and
$\alpha_s^{\text{\sm(5)}}(M_Z) = 0.1184$.

\begin{figure}[tbh!]
  \centering
  \subfloat[][]{%
    \includegraphics[width=0.49\textwidth]{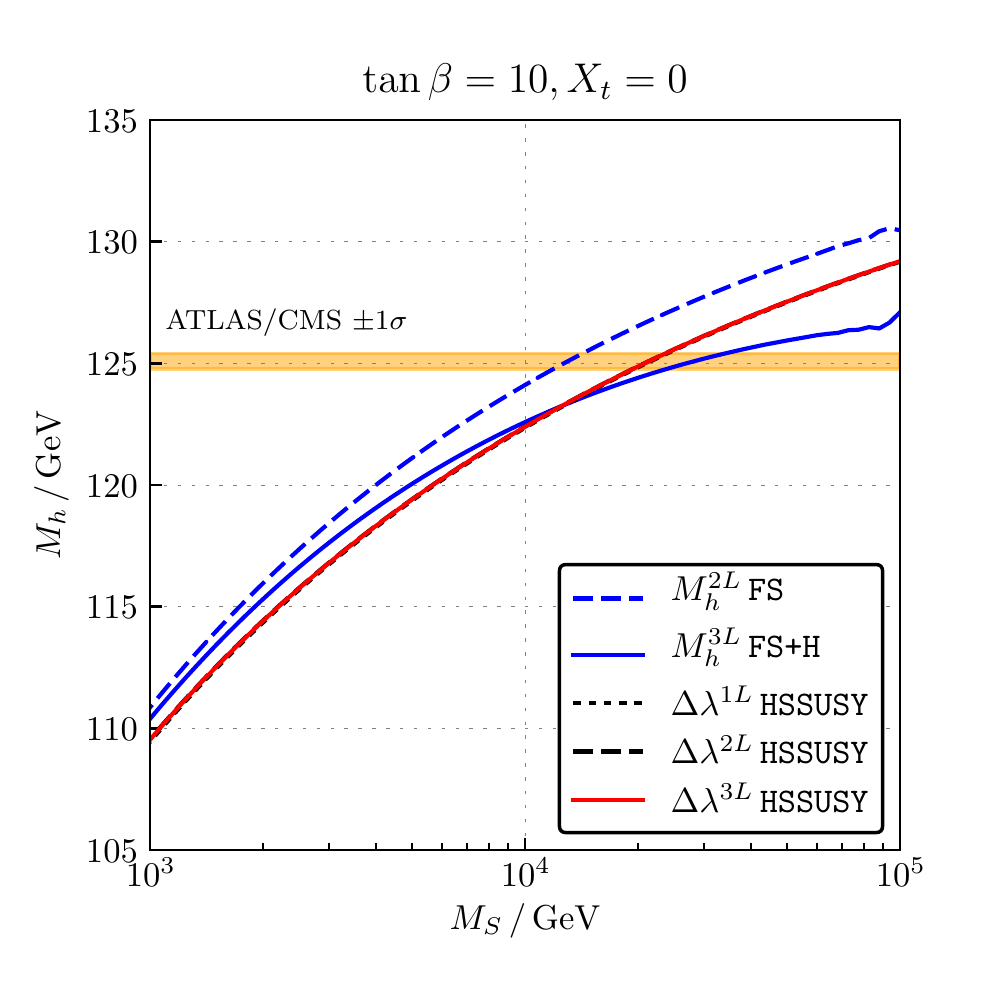}%
    \label{fig:Mh_abs_Xt-0}%
  }\hfill
  \subfloat[][]{%
    \includegraphics[width=0.49\textwidth]{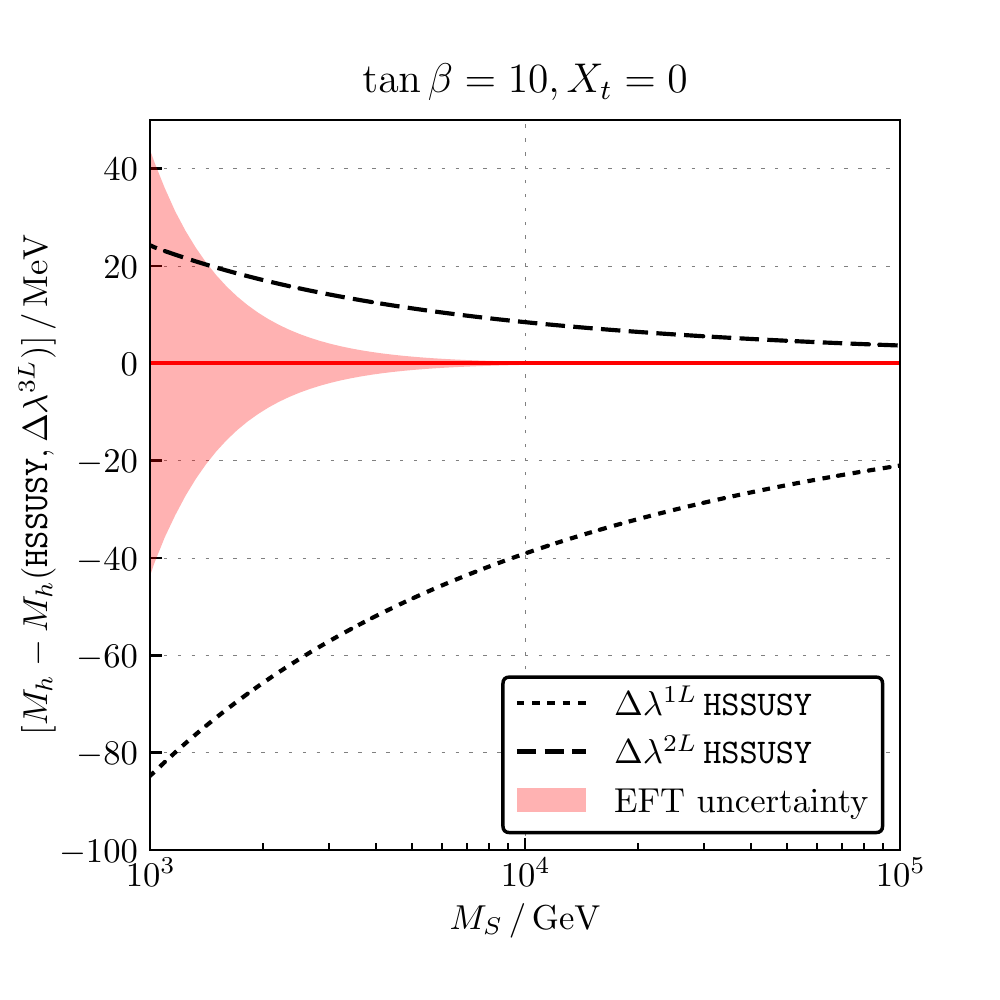}%
    \label{fig:Mh_diff_Xt-0}%
  }\\
  \subfloat[][]{%
    \includegraphics[width=0.49\textwidth]{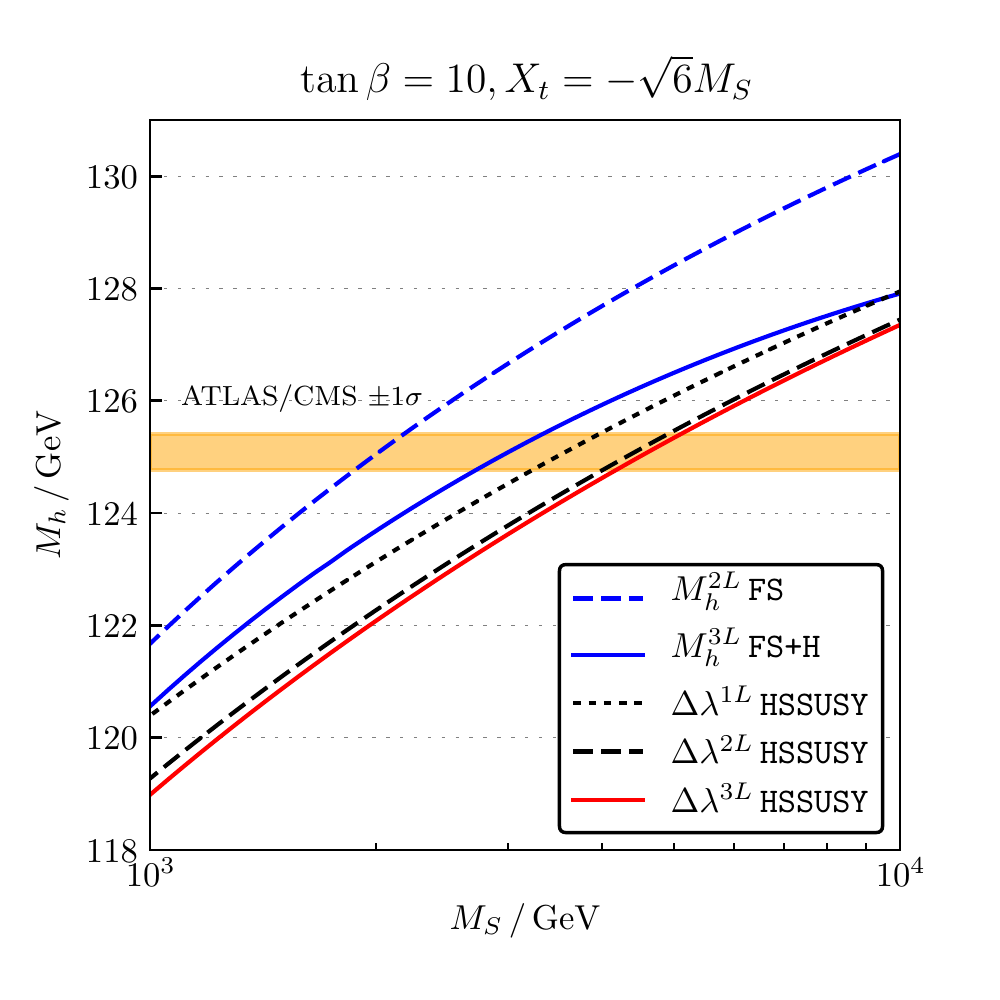}%
    \label{fig:Mh_abs_Xt--sqrt6}%
  }\hfill
  \subfloat[][]{%
    \includegraphics[width=0.49\textwidth]{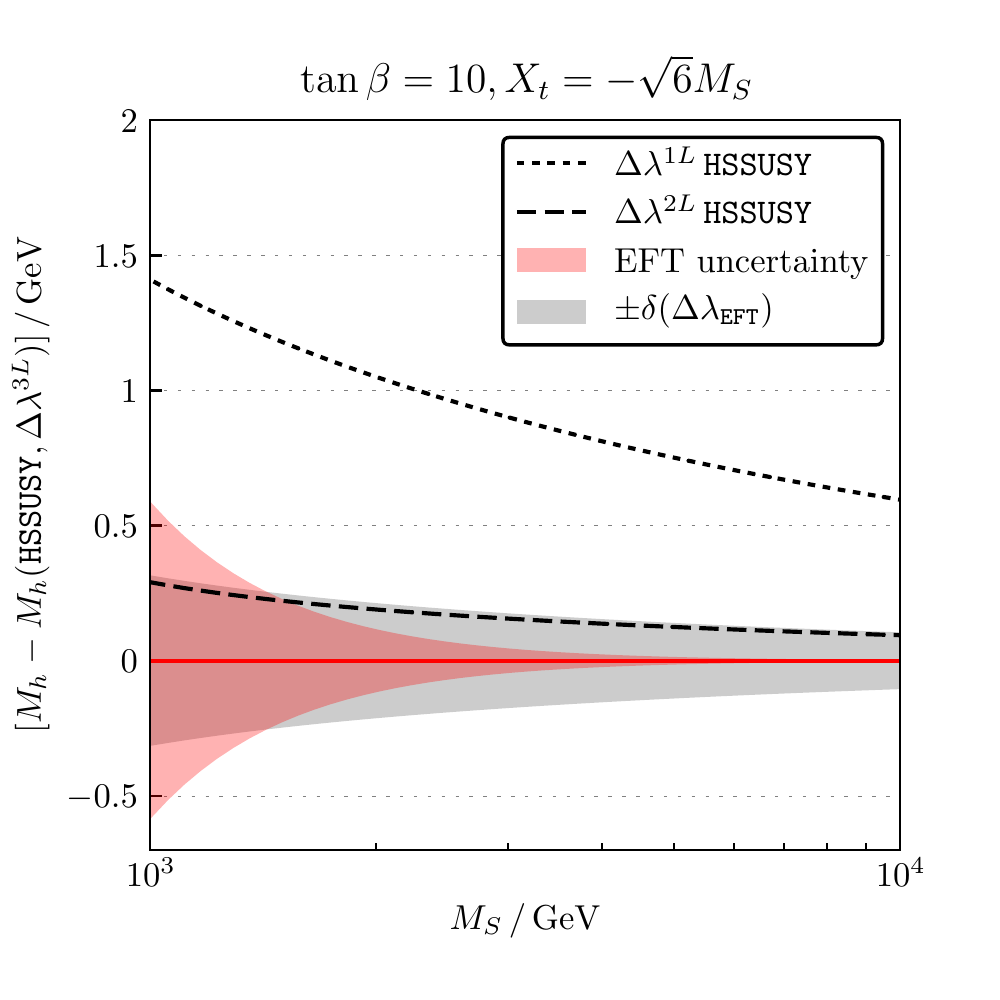}%
    \label{fig:Mh_diff_Xt--sqrt6}%
  }
  \caption{Comparison of the \three-loop \HSSUSY\ (\eft) calculation
    with lower order \eft\ and fixed-order \mssm\ calculations from
    the \code{FlexibleSUSY} package as a function of the \susy\ scale.
    The orange band marks the experimentally measured value of the
    Higgs mass as written in \eqn{eq:mhval}.}
  \label{fig:scan_MS}
\end{figure}
In \fig{fig:scan_MS} the effect of $\DlamSMEFT$ on the pure
\eft\ calculation of \HSSUSY\ is shown as a function of the \susy\ scale
$\ms$ for degenerate soft-breaking mass parameters, all set equal to
$\ms$. Furthermore, we set $\mu(\mus) = m_A(\mus) = \mus$,
$\tan\beta(\mus) = 10$, $A_t=X_t+\mu/\tan\beta$, while all other
trilinear couplings are set to zero.  The upper row shows a scenario with
vanishing stop mixing, $X_t(\mus) = 0$, the lower row shows one with
maximal stop mixing, $X_t(\mus) = -\sqrt{6}\ms$.
The left column of \fig{fig:scan_MS} displays the value of the
calculated Higgs boson mass for these two scenarios.  The blue dashed
line and the blue solid line show the \two- and \three-loop fixed-order
calculations of \FlexibleSUSY\ 2.1.0 and \FlexibleSUSY\ 2.1.0+\himalaya\ 2.0.1,
respectively.  The black dotted, dashed, and red solid line depict the
\eft\ calculations of \HSSUSY\ with $\lambdabar(\mus)$ calculated at the
\one-, \two-, and \three-loop level, respectively.  Here,
$\Delta\lambda^{1L}$ and $\Delta\lambda^{2L}$ denote all available
one- and two-loop corrections, respectively, and
$\Delta\lambda^{3L} = \DlamSMEFT$.  For comparison, the
yellow horizontal band shows the current experimental value for the
Higgs mass, see \eqn{eq:mhval}.  As was already observed for example
in \citeres{Athron:2016fuq,Athron:2017fvs,Allanach:2018fif}, we find
that in the range $\ms \geq 1\TeV$ the fixed-order and the \eft\
calculations deviate by several GeV.  This is to be expected, because
the \eft\ calculation resums the large logarithmic corrections (in
contrast to the fixed-order calculation) and above $\ms \gtrsim 1\TeV$
the neglected terms of $\order{\vbar^2/\ms^2}$ are negligible
\cite{Bahl:2017aev,Athron:2017fvs,Staub:2017jnp}.

As the black dashed and solid red line are hardly distinguishable in these
plots, we show the shift relative to the \one- and \two-loop
calculations of \HSSUSY\ in the right column of \fig{fig:scan_MS}. The
gray band in \fig{fig:Mh_diff_Xt--sqrt6} corresponds to the theoretical
uncertainty on the result due to the hierarchy expansions of the
\htm\ result, evaluated according to
\eqn{eq:delta_lambda_uncertainty_eft}; it amounts to more than 100\% of
the central shift for maximal mixing. For $X_t=0$, this uncertainty is
zero, see \eqn{eq:xt_uncertainty_deg}, because we also set
$\mus=\ms$. This is consistent with the fact that in this case, the
degenerate-mass limit of the \htm\ result is exact. The red band shows
the ``\eft\ uncertainty'' as defined in
\citeres{Bagnaschi:2014rsa,Vega:2015fna,Allanach:2018fif}, estimating
effects from missing terms of $\order{\vbar^2/\ms^2}$.  We see that the
impact of \DlamSMEFT\ is largely negative with respect to the \two-loop
threshold correction, $\Delta\lambda^{2L}$, and may reduce the Higgs
mass by up to $0.6\GeV$ for maximal mixing when considering all values
in the grey uncertainty band.  For zero stop mixing, the shift is
significantly smaller ($\lesssim 20\MeV$). 

\begin{figure}[tbh]
  \centering
  \includegraphics[width=0.49\textwidth]{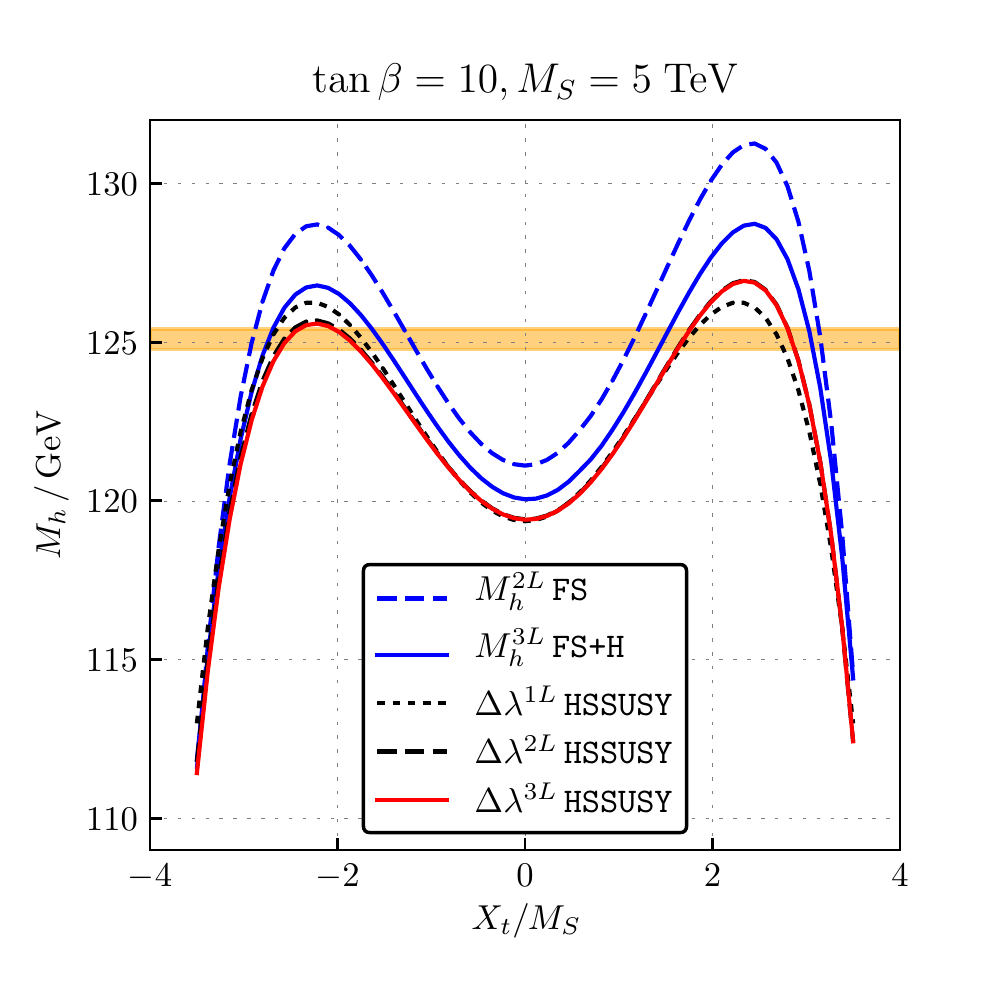}\hfill
  \includegraphics[width=0.49\textwidth]{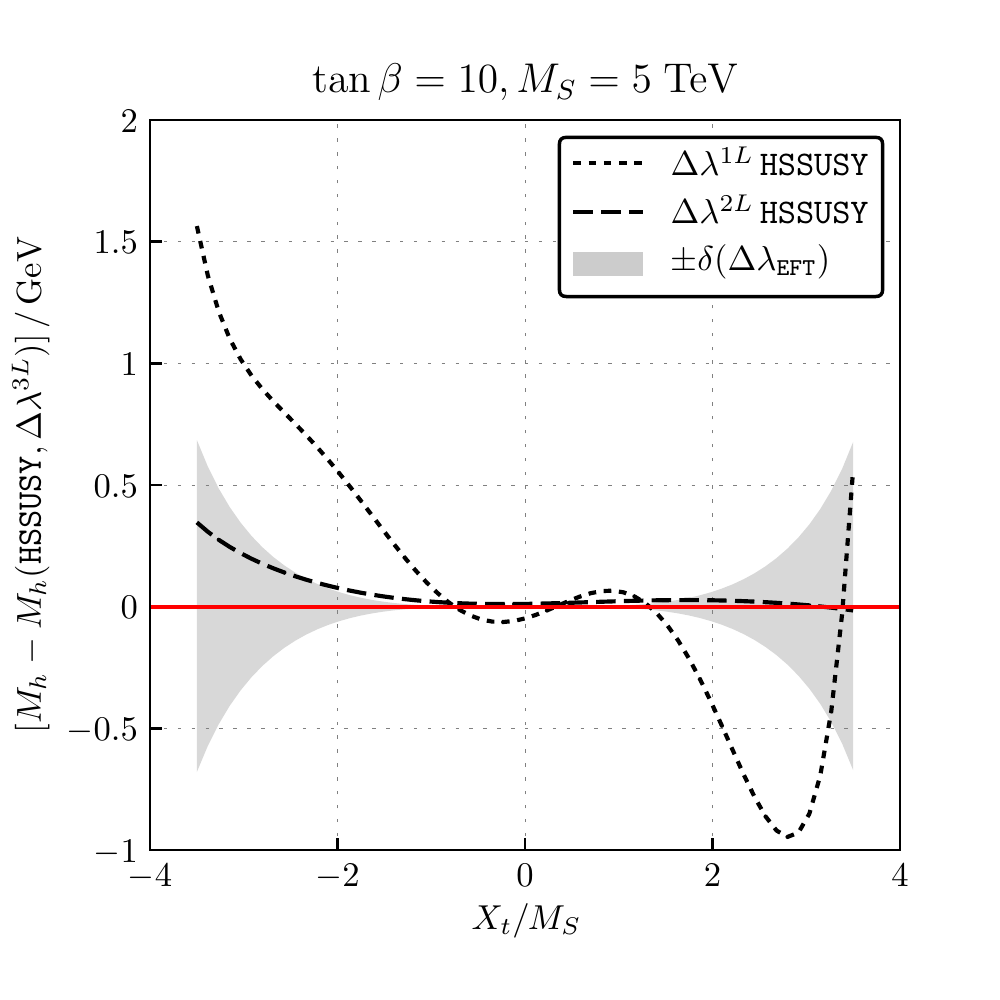}
  \caption{Comparison of the \three-loop \HSSUSY\ (\eft) calculation
    with lower order \eft\ and fixed-order \mssm\ calculations from
    the \code{FlexibleSUSY} package as a function of the relative stop
    mixing.}
  \label{fig:scan_xt}
\end{figure}
In \fig{fig:scan_xt}, the Higgs mass prediction is shown as a function
of the relative stop mixing parameter $x_t=X_t/\ms$ for a scenario with
$\tan\beta=10$ and $\ms = 5\TeV$, where both the fixed-order and the
\eft\ approach can accommodate for the experimentally observed value of
$\mhiggs$, \eqn{eq:mhval}, as long as $|x_t|$ is sufficiently large.
The right panel shows again the difference of the \three-loop
calculation of \HSSUSY\ with respect to the \one- and \two-loop
calculations.  In accordance with \fig{fig:scan_MS}, we find that the
shift induced by including \DlamSMEFT\ is negative by trend, and below
about $200\MeV$ for $x_t>-2$. Below that value, the effects could be of
order $1\GeV$, but the uncertainty of our approximation grows to about
100\% in this case, because the $x_t^4$ term is not included in
the hierarchy expansion of the \htm\ result for this scenario.

To get an idea of the maximal effect that $\DlamSMEFT$ can have on the
Higgs mass prediction, the blue band of \fig{fig:Mh_variation} shows the
variation of $\mhiggs$ when the \susy\ mass parameters $m_{Q,3}$,
$m_{U,3}$, $m_{D,3}$, and $\mgluino$ are varied simultaneously and independently within the interval
$[\ms/\sqrt{2},\sqrt{2}\ms]$ as a function of $\ms$, including the
uncertainty $\delta(\DlamSMEFT)$.\footnote{The choice of the interval
  $[\ms/\sqrt{2},\sqrt{2}\ms]$ ensures that for all scanned points there
  exists a suitable mass hierarchy which fits the parameter point with a
  moderate uncertainty $\DlamSMEFT$.  In the scanned parameter region,
  the most frequently chosen hierarchy is \code{h3} or one of its
  sub-hierarchies.}  The hatched region marks the range of \susy\ scales
where the lightest running stop mass is below $1\TeV$ for at least one
of the scanned points; in this case, the \eft\ may not be applicable.
For zero stop mixing (left panel), we find that $\DlamSMEFT$ can have an
effect up to $\approx - 150\MeV$ for $\ms\geq 1\TeV$.  In the region
where $\mstop{1} > 1\TeV$, the correction reduces to $-130\MeV$ at most.
The \three-loop correction decreases for larger \susy\ scales, mainly
due to the fact that the \sm\ couplings become smaller.  For maximal
stop mixing, $x_t = -\sqrt{6}$, the effect of the \three-loop correction
is significantly larger, and can reach $-1.25\GeV$ for $\mstop{1}
\gtrsim 1\TeV$.  The correction becomes particularly large when the
soft-breaking stop-mass parameters $m_{Q,3}$ and $m_{U,3}$ become small.
\begin{figure}[tbh]
  \centering
  \includegraphics[width=0.49\textwidth]{{{plots/lambda3L_size/scan_DMh_MS_TB-10_Xt-0}}}\hfill
  \includegraphics[width=0.49\textwidth]{{{plots/lambda3L_size/scan_DMh_MS_TB-10_Xt--2.44949}}}
  \caption{Variation of $\mhiggs$ when the \susy\ mass parameters are
    varied within the interval $[\ms/\sqrt{2},\sqrt{2}\ms]$ in \HSSUSY\ for
    $\tan\beta = 10$.  The left panel shows $X_t = 0$ and the right
    panel $X_t = -\sqrt{6}\ms$.  The blue band shows the maximal
    variation of $\mhiggs$ when the \three-loop correction
    $\DlamSMEFT \pm \delta(\DlamSMEFT)$ is included, with respect to the \two-loop
    calculation.  In the hatched region there is
    $\mstop{1}(\ms) \leq 1\TeV$ for at least one of the scanned
    parameter points.}
  \label{fig:Mh_variation}
\end{figure}

\section{Conclusions}

We have calculated the light \CP-even Higgs mass of the \mssm\ by
including all known fixed-order radiative corrections through
$\order{\alpha_t^2\alpha_s^2}$, and resumming the logarithmically
enhanced terms for a heavy \susy\ spectrum through fourth logarithmic
order in \susy\ \qcd. The only ingredient entering this result that was
unavailable in the literature up to now was the three-loop matching
coefficient at $\order{\alpha_t^2\alpha_s^2}$ for the quartic Higgs
coupling from the \sm\ to the \mssm. We derived it from the known
\three-loop corrections to the light \CP-even Higgs boson mass of
\citeres{Harlander:2008ju,Kant:2010tf}.  The coefficient is provided
both in terms of \drbarprime\ and \msbar\ parameters through its
implementation into the public \himalaya\ library, version 2.0.1. This
should facilitate its inclusion into spectrum generators which implement
the \eft\ approach. An uncertainty estimate is provided to account for
missing higher order terms in the mass hierarchy expansions.

Implementing $\DlamMSSM$ through \himalaya~2.0.1 into \HSSUSY, our
numerical analysis shows that the \three-loop correction tends to be
negative and may decrease the predicted Higgs boson pole mass by up to
$0.6\GeV$ for maximal stop mixing.  In scenarios with zero stop mixing,
the shift is significantly smaller, dropping to about $-25\MeV$ for
\susy\ mass parameters of around $1\TeV$.  For non-degenerate spectra
with $\mstop{1} \gtrsim 1\TeV$, the \three-loop correction can be of
the same size and reach up to $-1.25\GeV$ for low stop masses in
scenarios where a suitable mass hierarchy exists.  In scenarios where
no such hierarchy exists the correction may be significantly larger,
accompanied by a large expansion uncertainty.

\section*{Acknowledgments}

We are grateful to Matthias Steinhauser and Luminita Mihaila for helpful
communication, to Alexander Bednyakov for help with the extraction of
the \two-loop matching relation $(\Delta\alpha_t)_{\as^2}$ from the
results of \citere{Bednyakov:2002sf,Bednyakov:2005kt},
to Thomas Kwasnitza for making the mixed \two-loop corrections of
$\order{\bar\alpha_t^n\bar\alpha_b^m}$ to the Higgs pole mass in the
\sm\ available through \FlexibleSUSY,
and to Pietro
Slavich for helpful comments on the manuscript. This research was
supported in part by the Mexican \abbrev{CONACYT}, the German
\abbrev{DFG} through grant HA 2990/6-1 and the Research Unit \textit{New
  Physics at the LHC} (FOR 2239).

\clearpage
\appendix

\section{Documentation of Himalaya 2.0.1}\label{app:himalaya}

In this section we summarize technical details concerning the new
functionality of \himalaya\ 2.0.1.

\paragraph{Changes in Himalaya 2.0.1}

In \himalaya~2.0.1, we made changes to the hierarchy selection and to
some \three-loop expressions which may affect the calculated Higgs mass
at \three-loop level.  We list all of these changes below.
\begin{itemize}
\item In \himalaya\ 1.0.1, all input parameters are assumed to be given
  in the ``\htm\ scheme'', see \sct{sec:3fo}, and the output is provided
  in the same scheme by default.  Since most \mssm\ spectrum generators
  use the \drbarprime\ scheme, we have changed the definition of the
  input and output accordingly: In \himalaya\ 2.0.1, all input
  parameters are assumed to be given in the \drbarprime\ scheme.  The
  output is provided in the \drbarprime\ scheme by default.  Shifts to
  other renormalization schemes (\htm, \mdrbarprime, \ldots) are
  provided separately by \himalaya.

\item There are parameter scenarios where none of the \htm\ hierarchies
  fits to the \susy\ mass spectrum. \htm\ as well as \himalaya\ 1.0.1
  used the \texttt{h3} hierarchy in these cases, despite the fact that
  it does actually not fit. It turns out that the requirement
  \begin{align}
    \mstop{2} &> 1.3\msquark \,, &
    \mgluino  &> 1.3\msquark
    \label{eq:hiercond}
  \end{align}
  is sufficient to avoid these scenarios. \himalaya\ 2.0.1 will
  therefore throw an exception if the conditions \eqref{eq:hiercond}
  are not met.
\item For the highest order in $(\msquark^2-\mstop{i}^2)$ in the
  hierarchy expansions of \htm, we found disagreement with the
  logarithmic terms of the \eft\ approach. We therefore discarded these
  orders completely (also the non-logarithmic terms) in
  \himalaya.
\end{itemize}

\paragraph{Input parameters.}

With \himalaya\ 2.0.1 we extend the input parameters struct to a
more general form. Its new form is summarized in the following
listing:
\begin{lstlisting}[language=C++]
typedef Eigen::Matrix<double,2,1> V2;
typedef Eigen::Matrix<double,3,3> RM33;

struct Parameters {
   // DR'-bar parameters
   double scale{};         // renormalization scale
   double mu{};            // mu parameter, convention of
                           // [Phys.Rept. 117 (1985) 75-263]
   double g1{};            // GUT-normalized gauge coupling g1, with
                           // gY = g1*Sqrt[3/5]
   double g2{};            // gauge coupling g2 of SU(2)
   double g3{};            // gauge coupling g3 of SU(3)
   double vd{};            // VEV of down Higgs, with
                           // v = Sqrt[vu^2 + vd^2] ~ 246 GeV
   double vu{};            // VEV of up Higgs, with
                           // v = Sqrt[vu^2 + vd^2] ~ 246 GeV
   RM33 mq2{RM33::Zero()}; // soft-breaking squared left-handed
                           // squark mass parameters
   RM33 md2{RM33::Zero()}; // soft-breaking squared right-handed
                           // down-squark mass parameters
   RM33 mu2{RM33::Zero()}; // soft-breaking squared right-handed
                           // up-squark mass parameters
   RM33 ml2{RM33::Zero()}; // soft-breaking squared left-handed
                           // slepton mass parameters
   RM33 me2{RM33::Zero()}; // soft-breaking squared right-handed
                           // slepton mass parameters
   RM33 Au{RM33::Zero()};  // trilinear up type squark-Higgs
                           // coupling matrix
   RM33 Ad{RM33::Zero()};  // trilinear down type squark-Higgs
                           // coupling matrix
   RM33 Ae{RM33::Zero()};  // trilinear electron type squark-Higgs
                           // coupling matrix
   RM33 Yu{RM33::Zero()};  // up-type yukawa coupling matrix
   RM33 Yd{RM33::Zero()};  // down-type yukawa coupling matrix
   RM33 Ye{RM33::Zero()};  // electron-type yukawa coupling matrix

   // DR'-bar masses
   double M1{};            // bino
   double M2{};            // wino
   double MG{};            // gluino
   double MW{NaN};         // W
   double MZ{NaN};         // Z
   double Mt{NaN};         // top-quark
   double Mb{NaN};         // down-quark
   double Mtau{NaN};       // tau lepton
   double MA{};            // CP-odd Higgs
   V2 MSt{NaN, NaN};       // stops
   V2 MSb{NaN, NaN};       // sbottoms

   // DR'-bar mixing angles
   double s2t{NaN};        // sine of 2 times the stop mixing angle
   double s2b{NaN};        // sine of 2 times the sbottom mixing angle
};
\end{lstlisting}
The parameters initialized to \code{NaN} are optional and will be
calculated internally if not set to a finite value by the user.  Note
that all input parameters are interpreted as running \mssm\ parameters
in the \drbarprime\ scheme at the renormalization scale \code{scale}.

\paragraph{Calling Himalaya at the C++ level.}

Since the input parameters and the output of \himalaya\ 2.0.1 are
always defined in the \drbarprime\ scheme, we have removed the
\mdrbar\ flag in the constructor of the \code{HierarchyCalculator}.
The following source code listing shows an example call of \himalaya\
2.0.1:
\begin{lstlisting}[language=C++]
// create a new parameter point
himalaya::parameters point;
point.scale = 1000.; // GeV
point.mu    = 1000.; // GeV
// fill remaining parameters ...

// create the calculator class
himalaya::HierarchyCalculator hc(point);

// calculate all three-loop corrections of O(at*as^2)
himalaya::HierarchyObject hoTop = hc.calculateDMh3L(false);
\end{lstlisting}
The \code{HierarchyCalculator} class takes the parameter point as the
only mandatory argument.  To calculate the \three-loop corrections to
the \CP-even Higgs mass matrix or to the quartic Higgs coupling
$\lambda$, one needs to call the \code{calculateDMh3L} member function
of the created \code{HierarchyCalculator} object.  The
\code{calculateDMh3L} function takes a boolean argument to calculate the
corrections of $\order{\alpha_t^2\as^2}$ (argument is \code{false}) or
$\order{\alpha_b^2\as^2}$ (argument is \code{true}) to the \CP-even
Higgs mass matrix.  The function returns a \code{HierarchyObject} which
contains the calculated \three-loop results.

To convert the \three-loop results to other renormalization schemes,
the \code{HierarchyObject} class provides new member functions which
return additive shifts from the \drbarprime\ to any other scheme.  The
new member functions are listed in the following sub-section.

The following source code listing represents a complete example which
illustrates how the \three-loop correction of
$\order{\alpha_t^2\as^2}$ to the \CP-even Higgs mass matrix and to the
quartic Higgs coupling can be calculated with \himalaya\ 2.0.1.
\begin{lstlisting}[language=C++]
#include "HierarchyCalculator.hpp"
#include <cmath>

himalaya::Parameters make_point(double MS, double xt, double tb)
{
   himalaya::Parameters pars;

   const double MS2 = MS*MS;
   const double Xt = xt*MS;
   const double beta = std::atan(tb);

   pars.scale = MS;
   pars.mu = MS;
   pars.g1 = 0.46;
   pars.g2 = 0.65;
   pars.g3 = 1.166;
   pars.vd = 246*std::cos(beta);
   pars.vu = 246*std::sin(beta);
   pars.mq2 << MS2, 0, 0,
               0, MS2, 0,
               0, 0, MS2;
   pars.md2 << MS2, 0, 0,
               0, MS2, 0,
               0, 0, MS2;
   pars.mu2 << MS2, 0, 0,
               0, MS2, 0,
               0, 0, MS2;
   pars.ml2 << MS2, 0, 0,
               0, MS2, 0,
               0, 0, MS2;
   pars.me2 << MS2, 0, 0,
               0, MS2, 0,
               0, 0, MS2;
   pars.Au(2,2) = Xt + pars.mu/tb;
   pars.Yu(2,2) = 0.862;
   pars.Yd(2,2) = 0.133;
   pars.Ye(2,2) = 0.101;
   pars.MA = MS;
   pars.M1 = MS;
   pars.M2 = MS;
   pars.MG = MS;

   return pars;
}

int main()
{
   // create parameter point
   const auto point = make_point(2000, std::sqrt(6.), 20);

   // create calculator object
   himalaya::HierarchyCalculator hc(point);

   // calculate 3-loop corrections O(at^2*as^2)
   const auto ho = hc.calculateDMh3L(false);

   // get 3-loop contribution to CP-even Higgs mass matrix
   const auto dMh_3L = ho.getDMh(3);

   // get 3-loop contribution to lambda
   const auto delta_lambda_3L = ho.getDLambda(3);

   // get uncertainty for 3-loop lambda
   const auto delta_lambda_3L_uncertainty = ho.getDLambdaUncertainty(3);
}
\end{lstlisting}

\paragraph{New member functions of \code{HierarchyObject}.}

Below we list all member functions of \code{HierarchyObject} that are
new in \himalaya\ 2.0.1.
\begin{description}[align=left]
\item[\code{getDMhDRbarPrimeToMDRbarPrimeShift()}] Returns the additive
  shift to convert the Higgs mass matrix from the \drbarprime\ scheme
  at \three-loop level to the $\mdrbarprime$ scheme.

\item[\code{getDMhDRbarPrimeToH3mShift()}] Returns the additive shift to
  convert the Higgs mass matrix from the \drbarprime\ scheme at
  \three-loop level to the \htm\ scheme. In matrix form, the shift is given
  by:
  \begingroup
  \allowdisplaybreaks
  \begin{align}
    \begin{split}
      \left(\Delta M_{h,11}^2\right)_{\htm \to \drbarprime}
      &=C\mu^2 X_t^2 \left\{\mstop{1}^4
        -2 \mstop{1}^2 \mstop{2}^2 \ln
        \left(\frac{\mstop{1}^2}{\mstop{2}^2}\right)
        -\mstop{2}^4\right\},
    \end{split}\\
    \begin{split}
      \left(\Delta M_{h,12}^2\right)_{\htm \to \drbarprime} & = C \mu X_t
             \Bigg\{-\mstop{1}^4 \left(A_t X_t
             +3 \mstop{2}^2\right)+2 A_t \mstop{1}^2
             \mstop{2}^2 X_t
             \ln \left(\frac{\mstop{1}^2}{\mstop{2}^2}\right)\\
             &\quad+A_t \mstop{2}^4 X_t+\mstop{1}^6+3
             \mstop{1}^2 \mstop{2}^4-\mstop{2}^6\Bigg\},
    \end{split}\\
    \begin{split}
      \left(\Delta M_{h,21}^2\right)_{\htm\to\drbarprime} &=
      \left(\Delta M_{h,12}^2\right)_{\htm\to\drbarprime},
    \end{split}\\
    \begin{split}
      \left(\Delta M_{h,22}^2\right)_{\htm\to\drbarprime}
      & = C \Bigg\{\Delta_{12}
      \bigg[\mstop{1}^2 \left(A_t^2 X_t^2+4 A_t \mstop{2}^2
      X_t-\mstop{2}^4\right)-\mstop{1}^4
      \left(2 A_t X_t+\mstop{2}^2\right)\\
      &\quad+\left(\mstop{2}^3-A_t \mstop{2}
      X_t\right)^2+\mstop{1}^6\bigg]
      -2 A_t^2 \mstop{1}^2 \mstop{2}^2 X_t^2 \ln
      \left(\frac{\mstop{1}^2}{\mstop{2}^2}\right)\Bigg\},
    \end{split}
  \end{align}%
  \endgroup
  with
  \begin{equation}
    C = -\frac{8\alpha_t^2\as^2v^2\sinb{2}}{\mstop{1}^2\mstop{2}^2\Delta_{12}^3}
    \Bigg\{-6 (l_{S{\tilde g}}+1) \mgluino^2+10 (l_{S{\tilde q}}+1)
      \msquark^2 + \sum_{i=1}^2 (1+l_{S{{\tilde t}}_i})\mstop{i}^2\Bigg\}.
  \end{equation}

\item[\code{getDLambda(int loops)}] Returns the correction to the
  matching relation of $\lambda$ at $n$-loop(s) including
  prefactors. $n$ can be $0,1,2,3$, where $n=3$ corresponds to
  $\DlamMSSMEFT$.

\item[\code{getDLambdaDRbarPrimeToMSbarShift(int loops)}] Returns the
  additive shift $\DlamShiftDRpToMS$ of
  \eqn{eq:shift_DRbarprime_to_MSbar}, which accounts for the effect of
  a parameter conversion in $\lambda$ at $n$-loop(s) from the
  \drbarprime\ to the \msbar\ scheme, including prefactors.  $n$ can
  be $0,1,2,3$, where $n=3$ corresponds to the shift for
  $\DlamMSSMEFT$.

\item[\code{getDLambdaUncertainty(int loops)}] For $\code{loops} = 3$
  the function returns the uncertainty $\delta(\DlamMSSMEFT)$
  according to \eqn{eq:delta_lambda_uncertainty_eft}, including the
  prefactors.  For $\code{loops} \neq 3$ the function returns zero.

\item[\code{getDMh2EFT(int loops)}] Returns
  $M_{h,\text{\eft},\code{<loops>}}^2$ according to
  \eqn{eq:Mh2_eft_cut_off} at $n$-loop(s). $n$ can be $0$, $1$, $2$,
  $3$, where $n=3$ includes the contribution of $\DlamMSSMEFT$.  The
  \three-loop result \code{getDMh2EFT(3)} can be used to extract
  $\DlamMSSM$ from an alternative fixed-order calculation,
  following the procedure introduced in this paper.  See below for an
  example.
\end{description}

\paragraph{Extracting $\DlamMSSM$ from alternative \three-loop
  calculations of the Higgs mass.}

The results for matching coefficient $\DlamMSSMEFT$ presented in this
paper rely on the \htm\ result for the \three-loop Higgs mass.  By using
the member functions \code{getDMh2EFT(int)} and \code{getDLambda(int)}
of the \code{HierarchyObject}, it is possible to extract the \three-loop
correction $\DlamMSSM$ from any other \three-loop fixed-order
\drbarprime\ $\order{\alpha_t^2\as^2}$ expression for the Higgs mass.
These two member functions return the following \three-loop
contributions
\begin{align}
  \code{getDMh2EFT(3)} &=
  M_{h,\text{\eft},3}^2\bigg|_{\DlamMSSM=0}
   + \kappa^3\vev^2 \alpha_t^2 \as^2\sinb{4} \DlamMSSMEFT - M_{h,\text{\eft},2}^2 \,, \\
  \code{getDLambda(3)} &= \kappa^3 \alpha_t^2 \as^2 \sinb{4} \DlamMSSMEFT \,.
\end{align}
with $M_{h,\text{\eft},n}^2$ and $\DlamMSSMEFT$ defined in
\sct{sec:extraction_of_lambda}.  By combining these functions with an
alternative \three-loop calculation $M^2_{h,\text{\eft},3}$ as
\begin{align}
  (\Delta\lambda_\text{alt})_{\alpha_t^2\as^2} \, v^2 &=
  M_{h,\text{\eft},3}^2 - M_{h,\text{\eft},2}^2
  - \code{getDMh2EFT(3)} + v^2 \times \code{getDLambda(3)} \\
  &= M^2_{h,\text{\eft},3} - M_{h,\text{\eft},3}^2(\mut,\mus)
  \bigg|_{\DlamMSSM=0}
\end{align}
one can extract the corresponding \three-loop correction
$(\Delta\lambda_\text{alt})_{\alpha_t^2\as^2}$.

\clearpage

\def\app#1#2#3{{\it Act.~Phys.~Pol.~}\jref{\bf B #1}{#2}{#3}}
\def\apa#1#2#3{{\it Act.~Phys.~Austr.~}\jref{\bf#1}{#2}{#3}}
\def\annphys#1#2#3{{\it Ann.~Phys.~}\jref{\bf #1}{#2}{#3}}
\def\cmp#1#2#3{{\it Comm.~Math.~Phys.~}\jref{\bf #1}{#2}{#3}}
\def\cpc#1#2#3{{\it Comp.~Phys.~Commun.~}\jref{\bf #1}{#2}{#3}}
\def\epjc#1#2#3{{\it Eur.\ Phys.\ J.\ }\jref{\bf C #1}{#2}{#3}}
\def\fortp#1#2#3{{\it Fortschr.~Phys.~}\jref{\bf#1}{#2}{#3}}
\def\ijmpc#1#2#3{{\it Int.~J.~Mod.~Phys.~}\jref{\bf C #1}{#2}{#3}}
\def\ijmpa#1#2#3{{\it Int.~J.~Mod.~Phys.~}\jref{\bf A #1}{#2}{#3}}
\def\jcp#1#2#3{{\it J.~Comp.~Phys.~}\jref{\bf #1}{#2}{#3}}
\def\jetp#1#2#3{{\it JETP~Lett.~}\jref{\bf #1}{#2}{#3}}
\def\jphysg#1#2#3{{\small\it J.~Phys.~G~}\jref{\bf #1}{#2}{#3}}
\def\jhep#1#2#3{{\small\it JHEP~}\jref{\bf #1}{#2}{#3}}
\def\mpl#1#2#3{{\it Mod.~Phys.~Lett.~}\jref{\bf A #1}{#2}{#3}}
\def\nima#1#2#3{{\it Nucl.~Inst.~Meth.~}\jref{\bf A #1}{#2}{#3}}
\def\npb#1#2#3{{\it Nucl.~Phys.~}\jref{\bf B #1}{#2}{#3}}
\def\nca#1#2#3{{\it Nuovo~Cim.~}\jref{\bf #1A}{#2}{#3}}
\def\plb#1#2#3{{\it Phys.~Lett.~}\jref{\bf B #1}{#2}{#3}}
\def\prc#1#2#3{{\it Phys.~Reports }\jref{\bf #1}{#2}{#3}}
\def\prd#1#2#3{{\it Phys.~Rev.~}\jref{\bf D #1}{#2}{#3}}
\def\pR#1#2#3{{\it Phys.~Rev.~}\jref{\bf #1}{#2}{#3}}
\def\prl#1#2#3{{\it Phys.~Rev.~Lett.~}\jref{\bf #1}{#2}{#3}}
\def\pr#1#2#3{{\it Phys.~Reports }\jref{\bf #1}{#2}{#3}}
\def\ptp#1#2#3{{\it Prog.~Theor.~Phys.~}\jref{\bf #1}{#2}{#3}}
\def\ppnp#1#2#3{{\it Prog.~Part.~Nucl.~Phys.~}\jref{\bf #1}{#2}{#3}}
\def\rmp#1#2#3{{\it Rev.~Mod.~Phys.~}\jref{\bf #1}{#2}{#3}}
\def\sovnp#1#2#3{{\it Sov.~J.~Nucl.~Phys.~}\jref{\bf #1}{#2}{#3}}
\def\sovus#1#2#3{{\it Sov.~Phys.~Usp.~}\jref{\bf #1}{#2}{#3}}
\def\tmf#1#2#3{{\it Teor.~Mat.~Fiz.~}\jref{\bf #1}{#2}{#3}}
\def\tmp#1#2#3{{\it Theor.~Math.~Phys.~}\jref{\bf #1}{#2}{#3}}
\def\yadfiz#1#2#3{{\it Yad.~Fiz.~}\jref{\bf #1}{#2}{#3}}
\def\zpc#1#2#3{{\it Z.~Phys.~}\jref{\bf C #1}{#2}{#3}}
\def\ibid#1#2#3{{ibid.~}\jref{\bf #1}{#2}{#3}}
\def\otherjournal#1#2#3#4{{\it #1}\jref{\bf #2}{#3}{#4}}
\newcommand{\jref}[3]{{\bf #1}, #3 (#2)}
\newcommand{\hepph}[1]{{\tt [hep-ph/#1]}}
\newcommand{\mathph}[1]{{\tt [math-ph/#1]}}
\newcommand{\arxiv}[2]{{\tt arXiv:#1}}
\newcommand{\bibentry}[4]{#1, {\it #2}, #3\ifthenelse{\equal{#4}{}}{}{, }#4.}

\end{document}